\pdfoutput=1 
\documentclass[useAMS,usenatbib]{mn2e}
\usepackage{graphicx}
\usepackage{amsmath}
\usepackage{txfonts}
\usepackage{amssymb}
\usepackage{subfig}
\makeatletter
\newcommand*{\rom}[1]{\expandafter\@slowromancap\romannumeral #1@}
\makeatother

\title[Host galaxies of luminous z$\sim$0.6 quasars]{Host galaxies of luminous z$\sim$0.6 quasars: Major mergers are not prevalent at the highest AGN luminosities \footnote{Based on observations made with the NASA/ESA Hubble Space Telescope, obtained from the data archive at the Space Telescope Science Institute. STScI is operated by the Association of Universities for Research in Astronomy, Inc. under NASA contract NAS 5-26555}}
\author[C. Villforth et al.]{C. Villforth$^{1,2}$, T. Hamilton$^{3}$, M. M. Pawlik$^{2}$, T. Hewlett$^{2}$, K. Rowlands$^{2}$,  H. Herbst$^{4}$, \newauthor F. Shankar$^{5}$, A. Fontana$^6$, F. Hamann$^{4,8}$, A. Koekemoer$^{7}$, J. Pforr$^{9, 10}$, J. Trump$^{11,12}$, S. Wuyts$^{1}$\\
$^{1}$ University of Bath, Department of Physics, Claverton Down, BA2 7AY, Bath, UK\\
$^{2}$ SUPA, University of St. Andrews, School of Physics and Astronomy, North Haugh, KY16 9SS, St Andrews, UK\\
$^{3}$ Shawnee State University, Department of Physics, 940 Second Street, Portsmouth Ohio 45662, United States\\
$^{4}$ Department of Astronomy, University of Florida, 32611 Gainesville, Florida, United States\\
$^{5}$ Department of Physics and Astronomy, University of Southampton, Highfield, SO17 1BJ, UK\\
$^{6}$ INAF-Osservatorio Astronomico di Roma, Via Frascati 33, Monte Porzio Catone, I-00040 Rome, Italy\\
$^{7}$ Space Telescope Science Institute, 3700 San Martin Drive, Baltimore, MD 21218\\
$^{8}$ Department of Physics \& Astronomy, University of California, Riverside, CA 92507, USA\\
$^{9}$ European Space Research and Technology Centre (ESA/ESTEC), Keplerlaan 1, 2201 AZ Noordwijk, The Netherlands \\
$^{10}$ Aix Marseille Université, CNRS, LAM (Laboratoire d'Astrophysique de Marseille) UMR 7326, 13388, Marseille, France
\\
$^{11}$ Department of Astronomy and Astrophysics and Institute for Gravitation and the Cosmos, The Pennsylvania State University,\\ 525 Davey Lab, University Park, PA 16802, USA\\
$^{12}$ Hubble Fellow\\
}
\begin{document}

\date{}

\pagerange{\pageref{firstpage}--\pageref{lastpage}} \pubyear{2016}

\maketitle

\label{firstpage}

\begin{abstract}
Galaxy interactions are thought to be one of the main triggers of Active Galactic Nuclei (AGN), especially at high luminosities, where the accreted gas mass during the AGN lifetime is substantial. Evidence for a connection between mergers and AGN, however, remains mixed. Possible triggering mechanisms remain particularly poorly understood for luminous AGN, which are thought to require triggering by major mergers, rather than secular processes. 
We analyse the host galaxies of a sample of 20 optically and X-ray selected luminous AGN (log($L_{bol}$ [erg/s]) $>$ 45) at z $\sim$ 0.6 using HST WFC3 data in the F160W/H band. 15/20 sources have resolved host galaxies. We create a control sample of mock AGN by matching the AGN host galaxies to a control sample of non-AGN galaxies.
Visual signs of disturbances are found in about 25\% of sources in both the AGN hosts and control galaxies. Using both visual classification and quantitative morphology measures, we show that the levels of disturbance are not enhanced when compared to a matched control sample.
We find no signs that major mergers play a dominant role in triggering AGN at high luminosities, suggesting that minor mergers and secular processes dominate AGN triggering up to the highest AGN luminosities. The upper limit on the enhanced fraction of major mergers is $\leqslant$20\%. While major mergers might increase the incidence of (luminous AGN), they are not the prevalent triggering mechanism in the population of unobscured AGN.
\end{abstract}


\section{Introduction}
\label{S:intro}

Since the discovery of Seyfert galaxies \citep{seyfert_nuclear_1943} and later on quasars \citep{schmidt_3c_1963}, we have come to understand that accretion onto super-massive black holes is the physical origin of Active Galactic Nuclei (AGN) over a wide range of luminosities and observed properties \citep{antonucci_unified_1993,urry_unified_1995} \footnote{In the remainder of the paper, we will use the term AGN to refer to accreting black holes of all luminosities}. It was later discovered that super-massive black holes are present also in the centres of inactive galaxies \citep[see][for a recent review]{kormendy_coevolution_2013}. The masses of those black holes are found to correlate with the properties of the host galaxies, such as the velocity dispersion, mass and absolute magnitude \citep[][and references therein]{novak_correlations_2006}. This correlation suggests a co-evolution, either in a stochastic manner \citep{peng_how_2007,jahnke_non-causal_2011} or through a direct causal connection. The causal connection is commonly suggested to be due to a co-evolution of galaxies and black holes during major mergers followed by negative feedback from the AGN quenching star formation in its host \citep[e.g.][]{sanders_ultraluminous_1988,di_matteo_energy_2005,hopkins_cosmological_2008}. There is however so far no clear evidence to support this picture.

Understanding the processes that grow black holes is therefore crucial for our understanding of galaxy evolution. In order to understand the growth of black holes, we have to understand which processes supply the gas for accretion into the central kpc of galaxies (fuelling) as well as the processes that drive accretion onto the black hole (triggering). Black hole accretion rates scale with luminosity, and range from $\dot{M} \sim 10^{-4} - 10\ M_{\odot}/yr$. AGN lifetimes have been estimated to be around $\sim 10^{7}-10^{9} yr$ using population estimates \citep{martini_quasar_2001}. This leads to total fuel masses for a single fuelling episode up to $\sim10^{8} M_{\odot}$ for typical quasar accretion rates for $L^*$ AGN. Although some studies suggests AGN "flickering" lifetimes as short as $\sim10^{5} yr$ \citep{king_agn_2015,hickox_black_2014,schawinski_active_2015}, those studies would still require overall fuelling cycles of $\sim 10^{7}-10^{9} yr$. AGN therefore require considerable gas masses to be fed to the centre of the galaxy on time-scale much shorter than the typical dynamical time-scale of the host galaxy.

Different physical processes might drive gas to the centres of galaxies, these include gas-rich major and minor mergers \citep[e.g.][]{di_matteo_energy_2005,shankar_black_2012,fontanot_dependence_2015}, bars \citep{shlosman_bars_1989,shankar_black_2012,fontanot_dependence_2015} as well as disk instabilities triggered by cold flows \citep[e.g.][]{dekel_cold_2009,bournaud_black_2011} at higher redshifts. Due to the different gas supply needs at different luminosities, it has been suggested that the fuelling mechanisms depend on the AGN luminosity, with lower luminosity AGN being fuelled by so called secular processes, such as bars or minor mergers, with major mergers being the dominant process at luminosities above $L^*$ \citep{hopkins_characteristic_2009,somerville_semi-analytic_2008,hopkins_we_2013}. Theoretical models reproduce galaxy luminosity functions assuming that mergers with a wide range of mass ratios fuel all AGN activity \citep[e.g.][]{shen_supermassive_2009,lapi_quasar_2006,wyithe_self-regulated_2003,shankar_constraints_2010}. In this paper, we focus on the question if major mergers dominate AGN triggering at the highest luminosities.

The properties of AGN host galaxies have been the subject of research over a considerable time-span. Early observations of local AGN host galaxies showed high incidences of major mergers \citep[e.g.][]{bahcall_hubble_1997,canalizo_quasi-stellar_2001}. Comparisons of local AGN host galaxies and starburst galaxies showed that the main association in the local universe is between mergers and starbursts rather than mergers and AGN \citep{veilleux_deep_2009}. Studies analysing the incidence of post-coalescence merger features in AGN host galaxies compared to control samples generally found no signs of enhanced merger features over a wide range of redshifts and luminosities \citep[e.g.][]{boehm_agn_2012,cisternas_bulk_2011,grogin_agn_2005,mechtley_most_2015,kocevski_candels:_2012,villforth_morphologies_2014}, although some studies find enhancements of merger features \citep{cotini_merger_2013,hong_correlation_2015}. In contrast, studies analysing heavily reddened or obscured AGN have found high incidences of merger features \citep{urrutia_evidence_2008,kocevski_are_2015,fan_most-luminous_2016}.  Studies analysing the incidence of AGN as a function of nearest neighbour separation find increased incidences of AGN in galaxies with close nearby neighbours \citep[e.g.][]{ellison_galaxy_2013,koss_merging_2010}. However, this effect disappears once the central star formation rate, tracing central gas densities, is corrected for \citep{sabater_triggering_2015}. The evidence for other triggering mechanisms such as bars or high-redshift clumpy disks is similarly mixed \citep[e.g.][]{cisternas_role_2014,trump_no_2014,bournaud_observed_2012}. 

Combining these different studies into a coherent view of AGN triggering is further complicated by the fact that different studies cover a wide range of redshifts and luminosities. A mix of redshifts will confuse the results since the incidence of different triggering processes might evolve with redshift. For example, the gas fraction evolves strongly with redshift \citep[e.g.][]{morokuma-matsui_redshift_2015,geach_evolution_2011,lagos_cosmic_2011,santini_evolution_2014,genzel_combined_2015}, affecting the efficiency of mergers driving gas to the central kpc as well as increasing the gas supply \citep[e.g.][]{lotz_effect_2010-1}. Similarly, the importance of cold flows in triggering disk instabilities shows a strong dependence on redshift \citep{cacciato_evolution_2012}.

In this study, we focus on analysing the properties of host galaxies of a sample of luminous AGN in a luminosity range in which galaxy interactions are expected to dominate the triggering of AGN. This study is a direct extension to a study of lower luminosity AGN by \citet{villforth_morphologies_2014}. We compare the host galaxy properties to a sample of control galaxies matched in absolute magnitude (tracing stellar mass) and redshift. The sample is introduced in Section \ref{S:sample}, observations and data reduction are presented in Section \ref{S:obs}, including the PSF construction in Section \ref{S:psf}, 2D fits in Section \ref{S:galfit}, the creation of a mock AGN control sample in Section \ref{S:control} and the quantitative and qualitative morphological analysis in Section \ref{S:morphology}. Results are presented in Section \ref{S:results}, followed by Discussion in Section \ref{S:discussion} and Conclusions (Section \ref{S:conclusions}). We use AB magnitudes throughout. The cosmology used is $H_{0}=70$(km/s)/Mpc, $\Omega_m=0.3$, $\Omega_{\lambda}$=0.7.

\section{Sample}
\label{S:sample}

The sample selection is aimed to compliment the lower-luminosity Chandra Deep Field South (CDFS) sample selected from the 4Ms catalogue \citep{xue_chandra_2011} analysed in \citet{villforth_morphologies_2014}. X-ray selection is therefore used to identify AGN in a uniform way over a wide range of luminosities.

The redshift range z=0.5-0.7 closely matches that in \citep{villforth_morphologies_2014}, who included sources up to z=0.8 to increase the statistical power of the study. The redshift is high enough to include a wide range of luminosities at an important cosmological epoch of galaxy formation, yet low enough to minimize surface brightness dimming. The redshift range chosen also spans only $\sim$1 Gyr in cosmic time, greatly limiting the influence of cosmic evolution.

\begin{figure}
\begin{center}
\includegraphics[width=8cm]{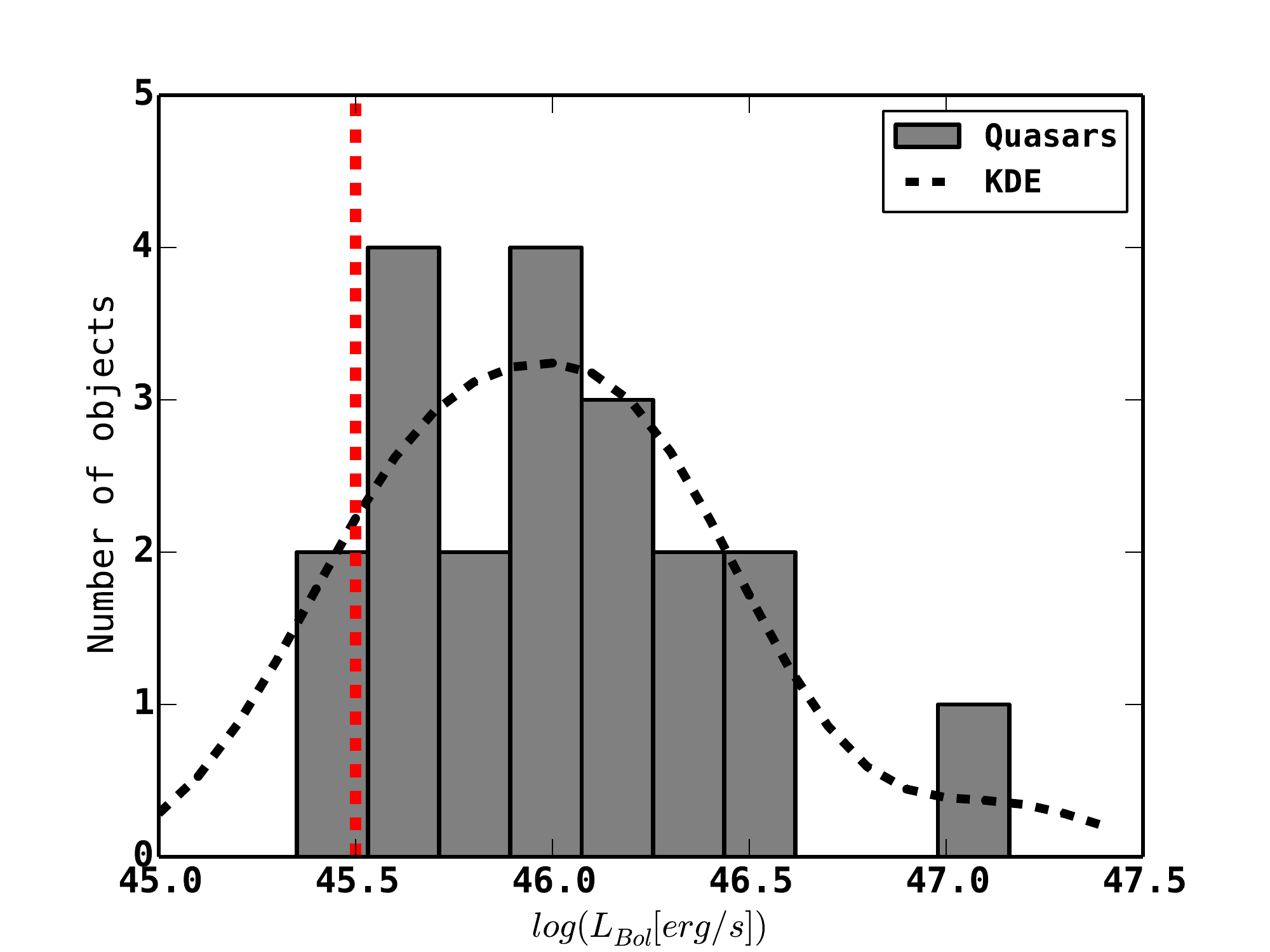}
\caption{Luminosity distribution of the bolometric luminosity for AGN in the sample. Bolometric luminosities are from \citet{shen_catalog_2011}. The black dashed line shows the Kernel Density Estimator (KDE) of the distribution, the vertical red dashed line shows the cut-off suggested by theoretical studies and discussed in the introduction and discussion \citep[e.g.][]{somerville_semi-analytic_2008}.}
\label{F:L_hist}
\end{center}
\end{figure}

We select 20 AGN from ROSAT-SDSS \citep{anderson_large_2007} with $L_{bol} > 10^{45} erg/s$, the theoretical cut-off for merger triggering assumed in a number of theoretical and semi-empirical models \citep{somerville_semi-analytic_2008} as well as the approximate upper limit of the CDFS sample from \citet{villforth_morphologies_2014}. We selected sources without nearby bright stars to avoid persistence or ghosting on the chip. The resulting luminosity distribution of the sample is shown in Fig. \ref{F:L_hist}. This selection method is inherently biased against heavily obscured sources. This selection effect is discussed in Section \ref{S:discussion}.

\begin{table}
\begin{minipage}{90mm}
\caption{Basic properties of the AGN in the sample. Name: Rosat SDSS ID; z: redshift; log(L$_{bol}$ [erg/s]): bolometric luminosity in erg/s from \citep{shen_catalog_2011}; log(M$_{BH}$ [M$_{\odot}$]): black hole mass from \citep{shen_catalog_2011}}
\begin{center}
\begin{tabular}{cccc}
\hline
ID & z & log(L$_{bol}$ [erg/s]) & log(M$_{BH}$ [M$_{\odot}$])\\ 
\hline
RASS 102 & 0.558 & 46.0 & 7.8 \\
RASS 126 & 0.660 & 46.2 & 9.1 \\
RASS 194 & 0.624 & 46.2 & 8.1 \\
RASS 236 & 0.700 & 46.2 & 8.5  \\
RASS 1345 & 0.512 & 45.6 & 8.4 \\
RASS 2019 & 0.676 & 45.7 & 8.7 \\
RASS 3845 & 0.533 & 45.4 & 8.8 \\
RASS 3970 & 0.515 & 46.0 & 8.7 \\
RASS 4050 & 0.600 & 46.3 & 8.7 \\
RASS 4339 & 0.666 & 45.6 & 8.8 \\
RASS 4406 & 0.627 & 46.6 & 10.0 \\
RASS 5169 & 0.671 & 45.7 & 9.2 \\
RASS 6044 & 0.681 & 46.3 & 8.8 \\
RASS 6057 & 0.600 & 47.2 & 9.1 \\
RASS 6107 & 0.590 & 46.5 & 8.8 \\
RASS 6126 & 0.530 & 45.7 & 8.8 \\
RASS 6133 & 0.638 & 46.0 & 8.8 \\
RASS 6135 & 0.620 & 46.0 & 9.5 \\
RASS 6156 & 0.688 & 45.8 & 9.1 \\ 
RASS 6193 & 0.611 & 45.4 & 9.3 \\
\hline
\end{tabular}
\end{center}
\label{T:sample}
\end{minipage}
\end{table}

\section{Observations and Data Analysis}
\label{S:obs}

Each of the 20 AGN selected was observed using the Hubble Space Telescope (HST) Wide Field Camera 3 (WFC3) in F160W (Broad H) for 2243s. This total exposure time was subdivided into nine exposures of 249s. Sub-pixel dithering with step sizes around 12" was performed between individual exposures to optimally sample the point spread function (PSF) and avoid bad regions on the chip. The log-linear STEP50 read-out pattern was used to perform several non-destructive reads of the central point source while sampling the faint extended emission using the linear steps. This ensured that the PSF stayed well constrained even when the central pixels of the PSF approached the non-linear regime of the chip during the exposure time. The central pixels of the AGN were checked for saturation and non-linearity. While pixel values in the non-linear regime where reached in some sources, the frequent samples in the beginning of the exposures yielded good sampling and therefore no problems in the reconstruction of the PSF in the central region. For details, see the WFC3 handbook\footnote{Dressel, L., 2016. “Wide Field Camera 3 Instrument Handbook, Version 8.0” (Baltimore: STScI)}.

All data were reduced using \textsc{AstroDrizzle} \citep{fruchter_betadrizzle:_2010, gonzaga_drizzlepac_2012}, the newest version of the \textsc{Drizzle} algorithm. The data were drizzled to a final frame with double the resolution of the initial images, resulting in a pixel scale of 0.0642"/pix, corresponding to approximately the Nyquist sampling of the data. Sampling to pixels smaller than the chip pixel size is possible due to the sub-pixel dithering performed. Smaller pixel sizes were tested but caused aliasing and problems with the PSF shapes, as also suggested in \citet{gonzaga_drizzlepac_2012}.

Additionally, we adjusted the drop size parameter for the drizzling algorithm. The drop size is the linear ratio between the size of the box used to drop the counts in a pixel onto the final drizzled pixels and the input pixel size \citep{gonzaga_drizzlepac_2012}. Large drop sizes effectively convolve the PSF with the pixel size, thereby decreasing resolution. In contrast, too small drop sizes result in only a small number of input frames contributing to a given pixel in the final drizzled image. Experimentation showed that drop sizes below 0.6 showed problems due to the limited number of dither positions, and a final pixfrac of 0.8 was chosen. We verified that the exact choice does not significantly affect the results. The chosen pixfrac optimised the detection of low surface brightness features.

Since data for each object were taken within a single orbit, the astrometry from the header was found to be sufficient for alignment between the different exposures and no re-alignment was performed.

\begin{figure*}
\begin{center}
\includegraphics[width=20cm]{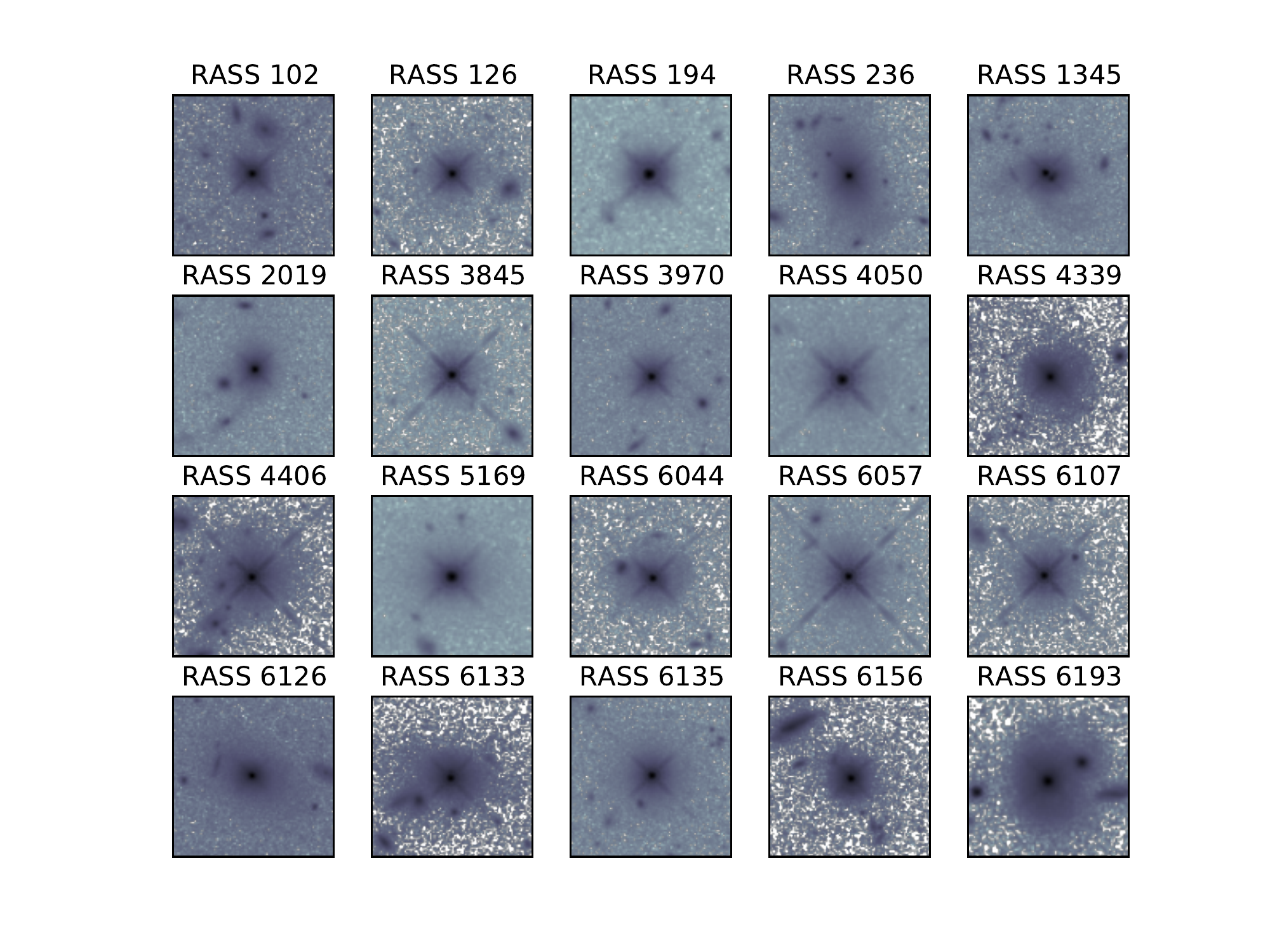}
\caption{HST F160W image cut-outs of the twenty AGN before 2D modelling. Individual cut-outs are 10\arcsec x 10\arcsec and are oriented as observed rather than North up and East left to show the similarity in PSF patterns.}
\label{F:gallery}
\end{center}
\end{figure*}

Image cut-outs showing all twenty AGN are shown in Figure \ref{F:gallery}. A wide range of visual morphologies are observed, including point-source dominated sources, clearly resolved host galaxies as well as disturbed sources. 

\subsection{PSF}
\label{S:psf}

The PSF was created using \textsc{starfit}\footnote{Hamilton, Timothy S., Starfit PSF Fitting Software (2014), ssucet.org/∼thamilton/research/starfit.html}, which creates and fits models of PSFs in HST images. It relies on the TinyTim \citep{krist_simulation_1995} software to make the basic PSF model, which \textsc{starfit} then fits to the point source, matching the PSF’s sub-pixel centring and telescope focus. These fits are performed for each individual sub-exposure of an orbit, and the final PSF is dithered following the same dither parameters used for the actual data (see Section \ref{S:obs}). This method not only includes the influences of focus changes during a single orbit, it also accounts for the effects of dithering on the PSF sampling.

Drizzled \textsc{starfit} PSFs were found to yield results consistent with those obtained from simple TinyTim PSF fits to single exposure data, showing that the drizzling process produced realistic PSFs. The overall fits are good, showing low residuals and allowing the recovery of host galaxies. Residuals for all fits are shown in the Appendix.

\subsection{Morphological Fits}
\label{S:galfit}

Morphological fits are performed using \textsc{galfit} \citep{peng_detailed_2002}. Each quasar is fit with a suite of different models to determine the best fit parameter. The results are presented in Table \ref{T:hostfit}.

In particular, each source was fit with a PSF and S\'{e}rsic profile. Multi-component fits using PSF + bulge + disk were tried, but did not improve the fits, as judged by the $\chi^2$ values returned by \textsc{GALFIT}. Most objects show well resolved host galaxies. In these cases, the host galaxy fits show physical values and no divergence, $\chi^2$ values show clear improvements compared to PSF only fits and visual comparison of PSF only and PSF+galaxy fits show clearly improved fits. Five sources are considered marginally detected since the fits diverged and alternative fits using disk or bulge components did not yield stable results (see Table \ref{T:hostfit}). The host galaxies in these cases are treated as undetected. The unresolved objects are still used in the visual inspection, but no control galaxies are matched for these objects. The effect of the unresolved host galaxies is discussed in Section  \ref{S:discussion}.

\begin{figure}
\begin{center}
\includegraphics[width=9cm]{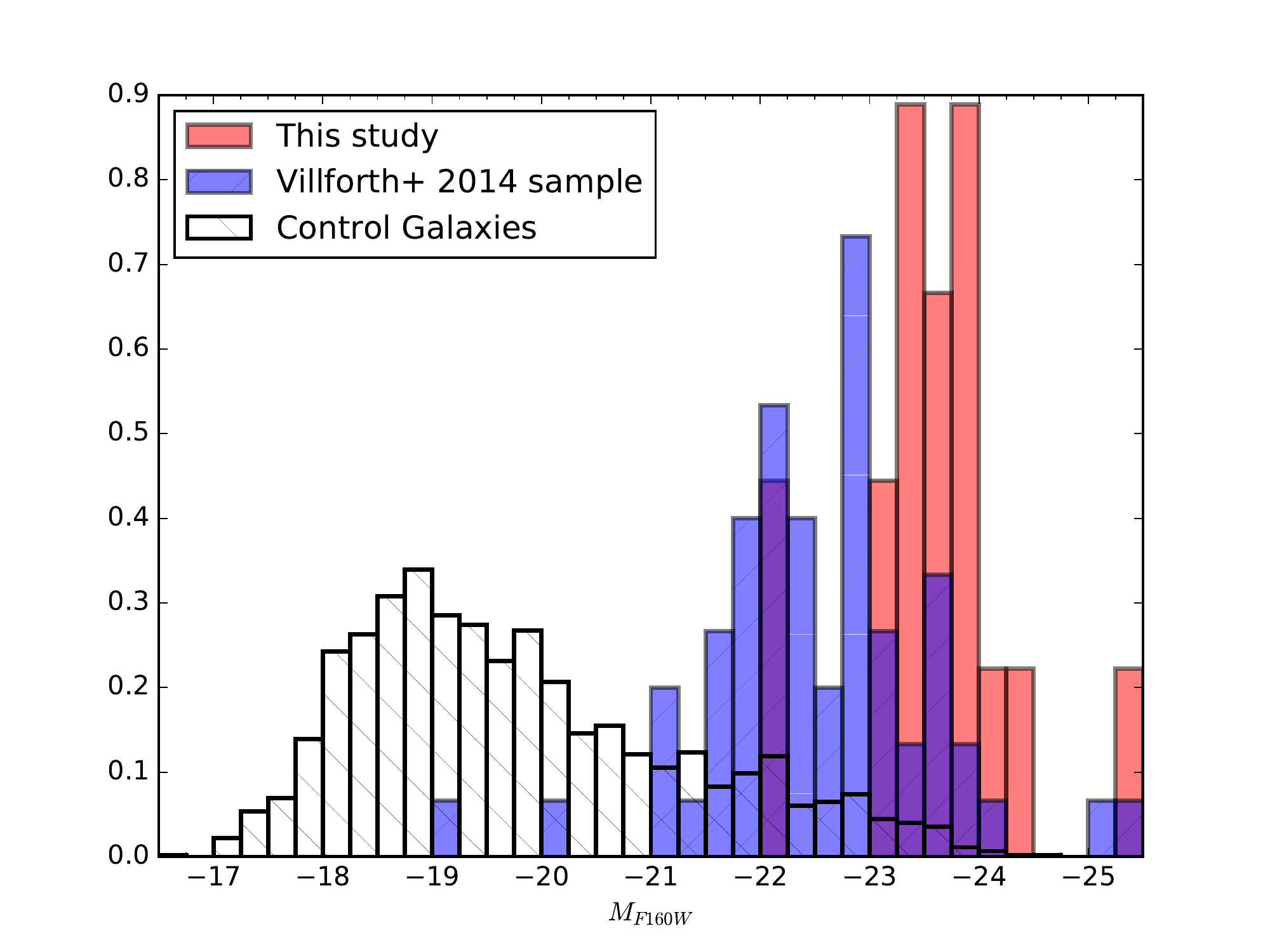}
\caption{Comparison of magnitude distributions in different samples. All histograms are normalized to allow comparison. Red: Histogram of host galaxy magnitudes for the best fit S\'{e}rsic model for all resolved host galaxies in the sample. Blue: distribution of host galaxy magnitudes for the lower luminosity sample from \citet{villforth_morphologies_2014}. Black hashed: full parent comparison sample (before matching) of inactive galaxies in the same redshift range in CANDELS GOODS-S.}
\label{F:galcomp}
\end{center}
\end{figure}

The detected host galaxies are luminous (see Fig. \ref{F:galcomp}), with absolute magnitudes $\sim$ -23.5 at rest-frame $\sim 1\mu$m. Following the findings in \citet{villforth_morphologies_2014}, who used stellar masses from \citep{santini_star_2009}, this implies stellar masses around $M_{*} \sim 10^{10-11} M_{\odot}$, making them massive galaxies.  We find reliable detections of galaxies up to $\sim$ 2.5 magnitudes fainter than the central point source. The sources with undetected host galaxies have considerable brighter point sources, so in these cases, the non-detections are consistent with the host galaxies being of similar absolute magnitude (and therefore stellar mass) to those in the resolved sample but are undetected due to stronger PSF contamination. The data are also consistent with fainter host galaxies in these cases. 

The S\'{e}rsic indices span a range, from $\sim 0.5-4.5$, with the majority of sources showing low S\'{e}rsic indices consistent with disk-like hosts. The host galaxies have effective radii between $\sim 0.4-4$", with the majority of sources having radii around 1", at the redshift of the sample, this corresponds to effective radii between 2.7-26 kpc, with the majority of sources having effective radii of $\sim 5-10$ kpc. We therefore find no indications that the objects in this sample are either unusually compact or show comparably high S\'{e}rsic indices, as would generally be expected for remnants of major gas-rich mergers.

\begin{table*}
\begin{minipage}{180mm}
\caption{Host galaxy properties for all AGN from \textsc{GALFIT} fits. All magnitudes are in observed F160W. ID, as used in table \ref{T:sample}; Resolved?: resolved flag, yes, or no; $m_{AGN}$: AGN magnitude; $m_{Galaxy}$: host galaxy magnitude; $r_{Galaxy}$ ["]  effective radius in arcseconds; S\'{e}rsic: S\'{e}rsic index; $M_{AGN}$ absolute AGN magnitude ; $M_{Galaxy}$ absolute galaxy magnitude}
\begin{center}
\begin{tabular}{cccccccc}
\hline
 & Resolved? & $m_{AGN}$ & $m_{Galaxy}$ & $r_{Galaxy}$ ["] & S\'{e}rsic & $M_{AGN}$ & $M_{Galaxy}$\\ 
\hline
RASS 102 & Y & 18.04 $\pm$ 0.01 & 19.07 $\pm$ 0.01 & 0.42 $\pm$ 0.01 & 1.45 $\pm$ 0.04 & -24.51 & -23.48 \\
RASS 126 & Y & 17.99 $\pm$ 0.02 & 19.19 $\pm$ 0.02 & 0.35 $\pm$ 0.01 & 2.39 $\pm$ 0.09 & -25.01 & -23.80 \\
RASS 194 & Y & 18.02 $\pm$ 0.01 & 19.62 $\pm$ 0.01 & 0.71 $\pm$ 0.01 & 0.47 $\pm$ 0.02 & -24.82 & -23.22 \\
RASS 236 & Y & 18.74 $\pm$ 0.04 & 17.88 $\pm$ 0.04 & 3.85 $\pm$ 0.22 & 4.48 $\pm$ 0.18 & -24.41 & -25.26 \\
RASS 1345 & Y & 18.37 $\pm$0.01 & 18.81 $\pm$ 0.01 & 0.65 $\pm$ 0.02 & 4.96 $\pm$ 0.13 & -23.95 & -23.51 \\
RASS 2019 & Y & 18.94 $\pm$ 0.01 & 19.34 $\pm$ 0.01 & 0.88 $\pm$ 0.01 & 1.68 $\pm$ 0.06 &  -24.12 & -23.71\\
RASS 3845 & N & - & - & - & - & - & - \\
RASS 3970 & Y & 17.80 $\pm$ 0.01 & 19.13 $\pm$ 0.01 & 0.53 $\pm$ 0.01 & 1.26 $\pm$ 0.03 & -24.54 & -23.20\\
RASS 4050 & N & - & - & - & - & - & - \\
RASS 4339 & Y & 19.58 $\pm$ 0.01 & 18.89 $\pm$ 0.01 & 0.68 $\pm$ 0.01 & 2.76 $\pm$ 0.08 &  -23.44 & -24.12\\
RASS 4406 & N & - & - & - & - & - & - \\
RASS 5169 & Y & 18.69 $\pm$ 0.02 & 18.77 $\pm$ 0.02 & 0.40 $\pm$ 0.01 & 4.48 $\pm$ 0.19 &  -24.35 & -24.27 \\
RASS 6044 & Y &  17.69 $\pm$ 0.02 & 19.74 $\pm$ 0.02 & 1.29 $\pm$ 0.03 & 1.52 $\pm$ 0.08 & -25.38 & -23.32 \\
RASS 6057 & N & - & - & - & - & - & - \\
RASS 6107 & N & - & - & - & - & - & - \\
RASS 6126 & Y & 20.00 $\pm$ 0.02 & 18.64 $\pm$ 0.02 & 1.78 $\pm$ 0.05 & 4.30 $\pm$ 0.14 &  -22.42 & -23.77 \\
RASS 6133 & Y & 17.80 $\pm$ 0.01 & 19.13 $\pm$ 0.01 & 0.92 $\pm$ 0.01 & 1.57 $\pm$ 0.05 & -25.10 & -23.76\\
RASS 6135 & Y & 17.34 $\pm$ 0.01 & 19.32 $\pm$ 0.01 & 0.98 $\pm$ 0.02 & 1.13 $\pm$ 0.05 & -25.47 & -23.51\\
RASS 6156 & Y &  18.73 $\pm$ 0.01 & 19.62 $\pm$ 0.01 & 0.81 $\pm$ 0.01 & 0.64 $\pm$ 0.02 & -24.37 &  -23.48\\ 
RASS 6193 & Y & 19.33 $\pm$ 0.02 & 18.81 $\pm$ 0.02 & 1.43 $\pm$ 0.03 & 1.60 $\pm$ 0.06 & -23.46 & -23.98\\
\hline
\end{tabular}
\end{center}
\label{T:hostfit}
\end{minipage}
\end{table*}

\subsection{Mock AGN Control Sample}
\label{S:control}

For the control sample, we use CANDELS data \citep{grogin_candels:_2011,koekemoer_candels:_2011} and catalogues from \citet{dahlen_detailed_2010} in GOODS-S, following \citet{villforth_morphologies_2014}. Control galaxies are matched to all resolved host galaxies using the $F160W$ host galaxy magnitude and redshift. \citet{villforth_morphologies_2014} showed that at these redshifts (z$\sim$0.6), where $F160W$ traces the rest-frame 1$\mu$m, the absolute magnitude is a good proxy for the stellar mass. Each quasar host galaxy is matched to five control galaxies, due to the fact that the AGN host galaxies are luminous and luminous galaxies are relatively rare, some control galaxies are matched to multiple AGN. 

To simulate the residuals caused by the bright central point source, a point source with an apparent magnitude matching that of the AGN is added to the centre of the matched control galaxies using model PSFs with realistic noise. The resulting mock AGN source is then fit using \textsc{Galfit} in the same manner as the main AGN sample. To avoid biasing the visual classification, the resulting images are randomized by rotating and mirroring them in a random fashion.

\subsection{Morphological analysis}
\label{S:morphology}

To determine merger rates, we use a combination of qualitative and quantitative measures. Visual inspection is performed by three human classifiers. Additionally, we use asymmetry measures \citep{conselice_symmetry_1997} as well as the updated shape asymmetry introduced by \citet{pawlik_shape_2016} which is shown to be more efficient at detecting faint extended emission such as tidal tails. Both quantitative and qualitative morphology measures will be discussed in the following sections.

\subsubsection{Visual Classification}
\label{S:visclas}

Visual classification of all twenty sources is performed by three human classifiers for comparison with quantitative morphology measures. In order to avoid biased classifications, classifiers are not involved in the morphological fits  and are therefore unable to distinguish between true AGN and control galaxies. Classification is performed on point source subtracted host galaxy images for both the AGN and control sample. The quasar and control sample are randomized to avoid bias by the classifiers. Classifiers are first asked to assign a general classification to each galaxy as follows:

\begin{itemize}
\item \textbf{N}: undisturbed
\item \textbf{D}: some signs of disturbance (asymmetric features)
\item \textbf{M}: clear merger (strong tidal features, double nuclei)
\item \textbf{X}: strong residual from point source subtraction, no classification possible
\end{itemize}

Additionally, classifiers are asked to provide flags for tidal features and the occurrence of nearby neighbours (within 10\arcsec). Neighbours do not need to show signs of interaction and since only one waveband is provided, the classifiers cannot distinguish between physical neighbours and foreground or background sources. The votes were consolidated by adopting the majority vote. In four cases, the overall vote was N, D, M, in which case D was adopted. In two cases, one classifier flagged the source for point source residuals, while the other two voters disagreed, in this case, the sources visual classification was set as X.

\subsubsection{Quantitative measures of disturbance}
\label{S:quantmorph}

Additional to the visual inspections, we chose to also use quantitative morphology measures to further determine the level of disturbance. As such, we use the asymmetry \citep[e.g.][]{schade_canada-france_1995,abraham_galaxy_1996,conselice_asymmetry_2000} as used in \citet[e.g.][]{villforth_morphologies_2014}. The asymmetry is defined as:

\begin{equation}
A \equiv \dfrac{ \sum |I_{0} - I_{180} | }{ 2 \sum | I_{0} |} 
\end{equation}
\noindent
where $I_{0}/I_{180}$ are the pixel fluxes of the original image and one rotated by 180$^{o}$ respectively. The central region of the galaxies are masked up to a radius of 5 pixels (0.300"), as the centre contains residuals from the PSF fitting. The exact radius was chosen from visual inspection of the frames.

Additionally, we use the Shape asymmetry ($A_{S}$) which is similar to the asymmetry described above, but is sensitive to different merger features, as described below. The shape asymmetry was introduced by \citet{pawlik_shape_2016} as an indicator of morphological rotational asymmetry of an object. It is measured by means of a detection mask - a binary image that contains information about the position of all pixels regarded as part of the object: all such pixels are assigned with a value of 1, while those that represent the sky are set to 0. The binary detection mask is computed by means of an 8-connected structure detection algorithm searching for all pixels within the original image with intensities above a given threshold, that are 8-connected to the central pixel (i.e. touching one of its edges or corners). 

For a robust detection of low surface-brightness features in the galaxy outskirts, the original images are passed through a $3\times 3$ running average (mean) filter and the threshold was set to 1$\sigma$ above the estimated sky background level. The binary detection masks obtained in such a way were then used to measure the shape asymmetry of the quasar host galaxies by considering the mathematical definition\footnote{The noise correction term used in the traditional definition was omitted in this case, as with the sky pixels set to 0, there is no random noise contribution to the measurement, and consequently, the obtained value of $A_{S}$ is independent of the choice of the aperture in which the measurement is performed.} of image asymmetry as defined above, but with $I_{0}$ being the pixel intensity in the \textit{detection} mask, and $I_{180}$, that in a rotated \textit{detection}n mask through a $180^{o}$ about the intensity-weighted minimum-asymmetry centroid. The position of the centroid was chosen out of a set of pixels accounting for the brightest $30\%$ the galaxy's total light, such as to minimise the corresponding value of the traditional asymmetry parameter, $A$, \citep{schade_canada-france_1995,abraham_galaxy_1996,conselice_asymmetry_2000} For more details on the computation of the shape asymmetry see \citet{pawlik_shape_2016}. Note that due to the binary nature of the images used for this method, a masking of central pixels is unnecessary.

\citep{pawlik_shape_2016} showed that the shape asymmetry performs better at detecting post-merger features such as tidal tails since the method weights areas with high and low flux equally. The shape asymmetry therefore down-weights central regions. While the two methods have overall similar scope, they are more sensitive to asymmetries in high and low flux regions, i.e. asymmetry would pick out mergers close to coalescence while shape asymmetry is more sensitive to later mergers.

\section{Results}
\label{S:results}

The majority of the host galaxies of luminous AGN are resolved, only five AGN have unresolved host galaxies, since the unresolved AGN are considerable brighter, the results are consistent with the host galaxies in the unresolved AGN being similar to the host galaxies found in the resolved AGN, although our data are also consistent with those sources hosting less luminous host galaxies. The AGN host galaxies are luminous massive galaxies, matching the most luminous galaxies found at the redshift of the sample (see Fig. \ref{F:galcomp}). The host galaxies of the current sample of luminous AGN are more luminous than the host galaxies of the lower luminosity sample studied by \citet{villforth_morphologies_2014} at the same redshift.

We discuss the morphologies of the AGN host galaxies and the mock control galaxies using both the visual classification (Section \ref{S:visclas}) and quantitative morphology measures (Section \ref{S:quantmorph}).

\begin{figure*}
\begin{center}
\includegraphics[width=8cm]{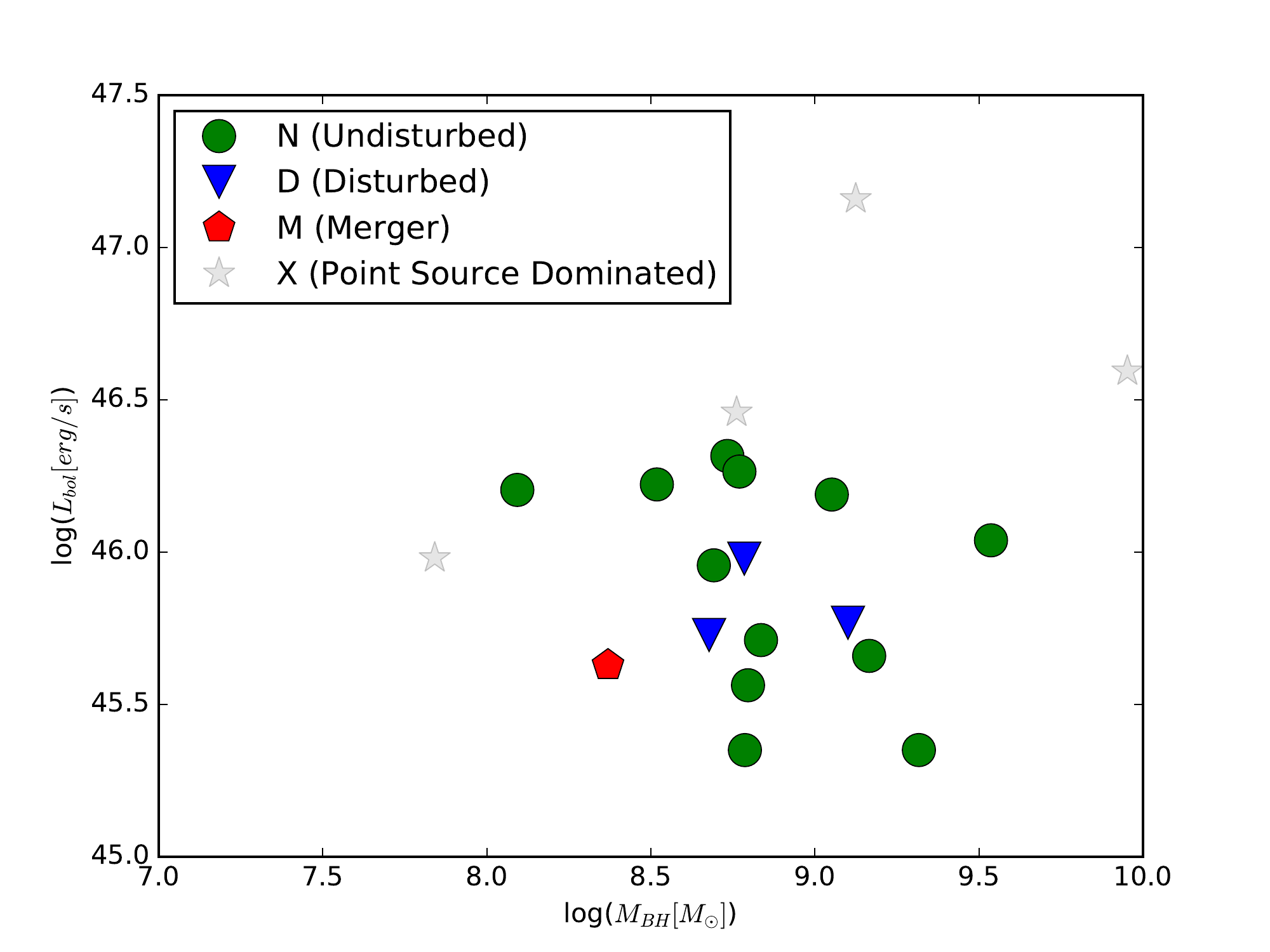}
\includegraphics[width=8cm]{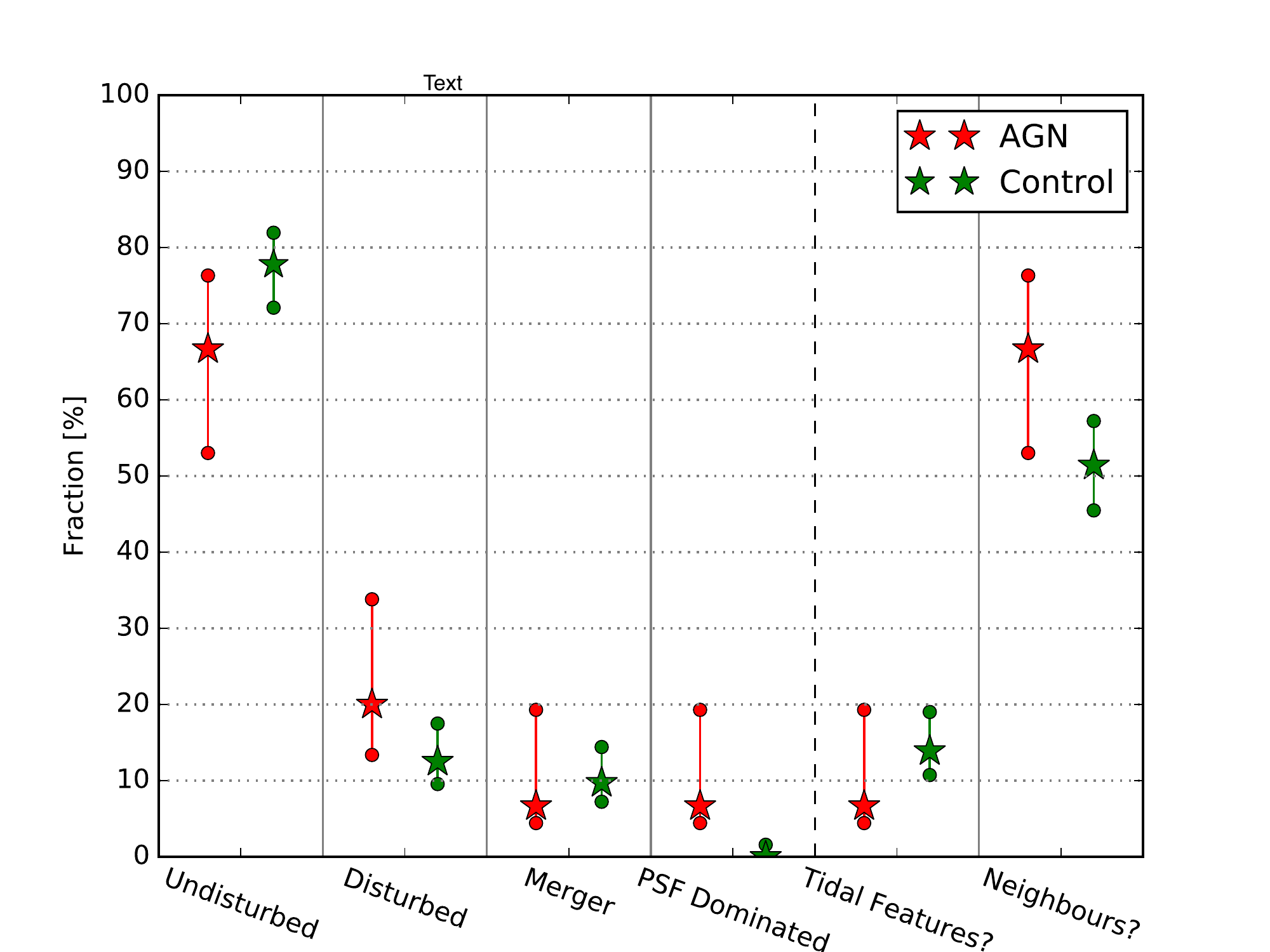}
\caption{Left: Visual classification of all resolved AGN host galaxies as a function of the bolometric luminosity and the black hole mass (see Table \ref{T:sample}). Right: Visual classification of all resolved AGN host galaxies and matched control galaxies. AGN are shown in red, control sample in green. The error bars show 1$\sigma$ confidence intervals calculated following \citet{cameron_estimation_2011}.}
\label{F:visclass}
\end{center}
\end{figure*}

The visual classification shows no signs of disturbance in the majority of cases, see Fig. \ref{F:visclass}. The most luminous AGN have too strong residuals for visual classification, the vast majority of sources are undisturbed, with four sources being classified as either disturbed or mergers. Only one source (RASS 1345, see Fig. \ref{F:gallery}) is classified as a merger, showing clear extended tidal tails. A comparison of the visual classification for the quasar host galaxies to the control galaxies shows that the fractions of disturbed galaxies is consistent with the morphologies found in the control samples (Fig. \ref{F:visclass}). While some minor differences are found in different morphological categories, none are statistically significant. Considering the size of the sample as well as the rate of objects classified as disturbed (merger + disturbed), the upper limit on excess disturbed sources in the AGN sample is 11\%/27\% at 3$\sigma$ (one-tailed p=0.05) for the merger and disturbed classification respectively (this is calculated from the $\beta$ statistics of the merger/disturbed fraction in the AGN sample, corrected for the control sample merger/disturbed fraction). We will discuss in section \ref{S:discussion} how incompleteness in the detection of mergers affects our results. To summarize, we find no statistically significant difference between the AGN and matched control sample in the visual classification.

\begin{figure*}
\begin{center}
\includegraphics[width=8cm]{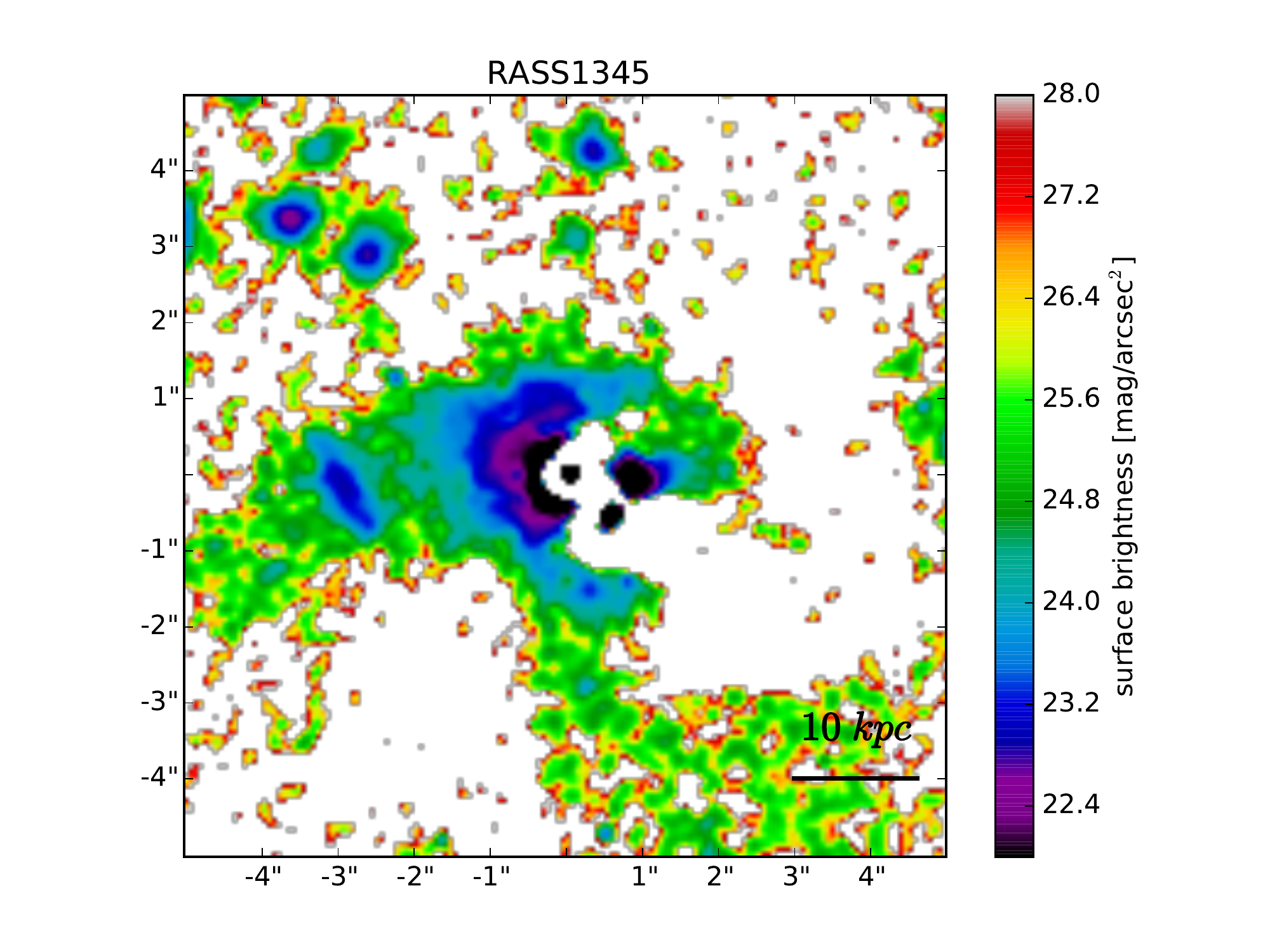}
\includegraphics[width=8cm]{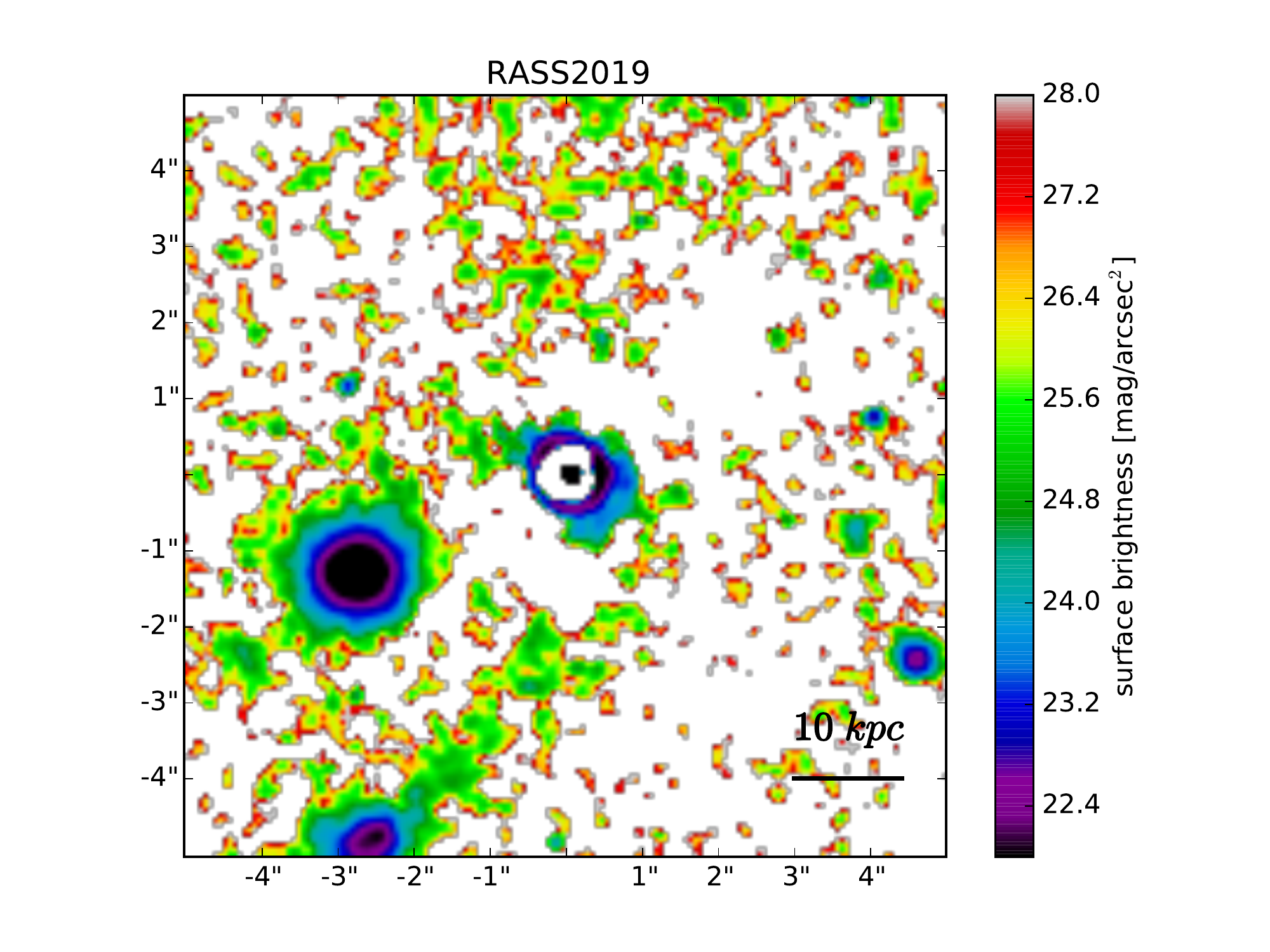}
\includegraphics[width=8cm]{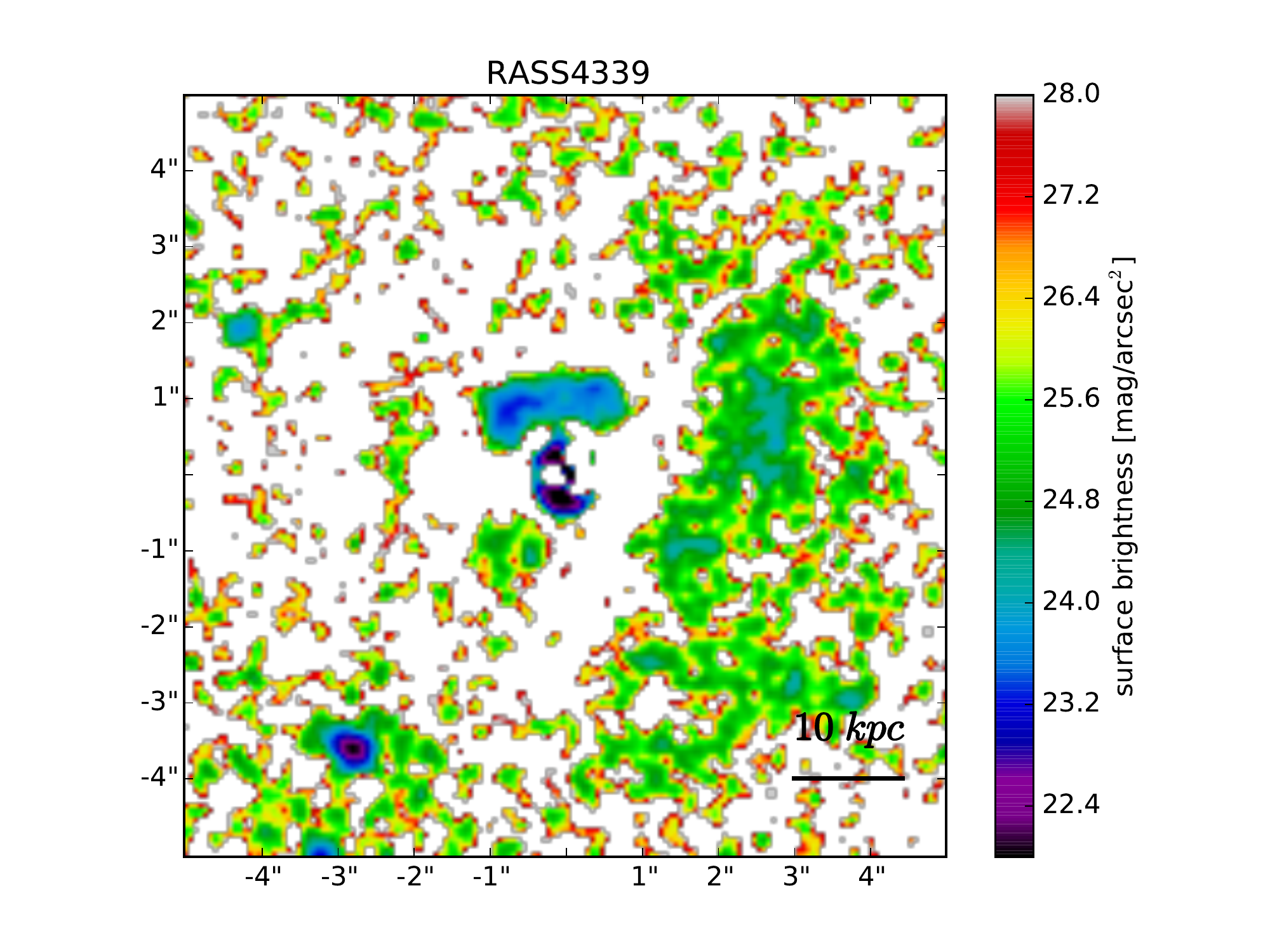}
\includegraphics[width=8cm]{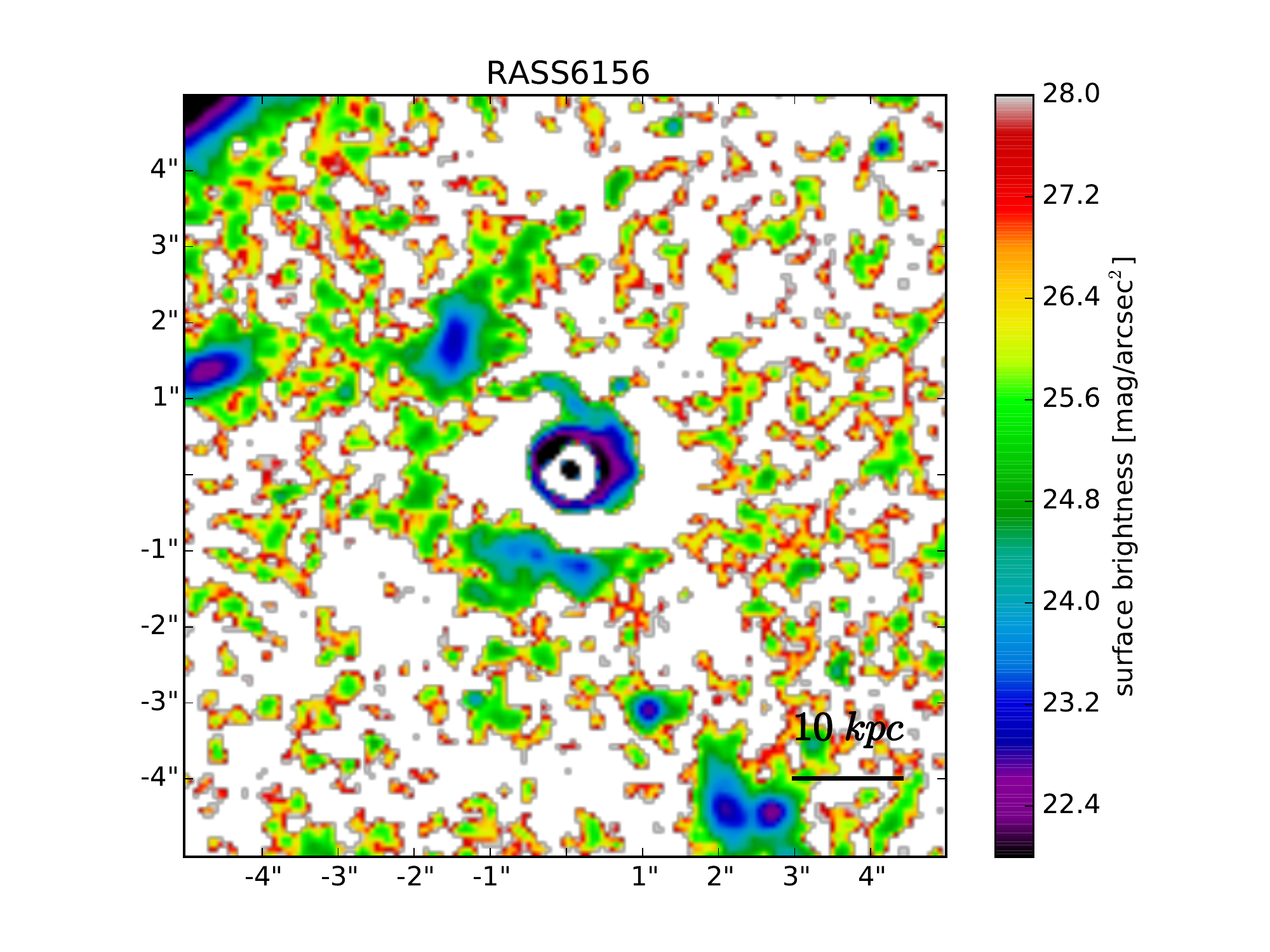}
\includegraphics[width=8cm]{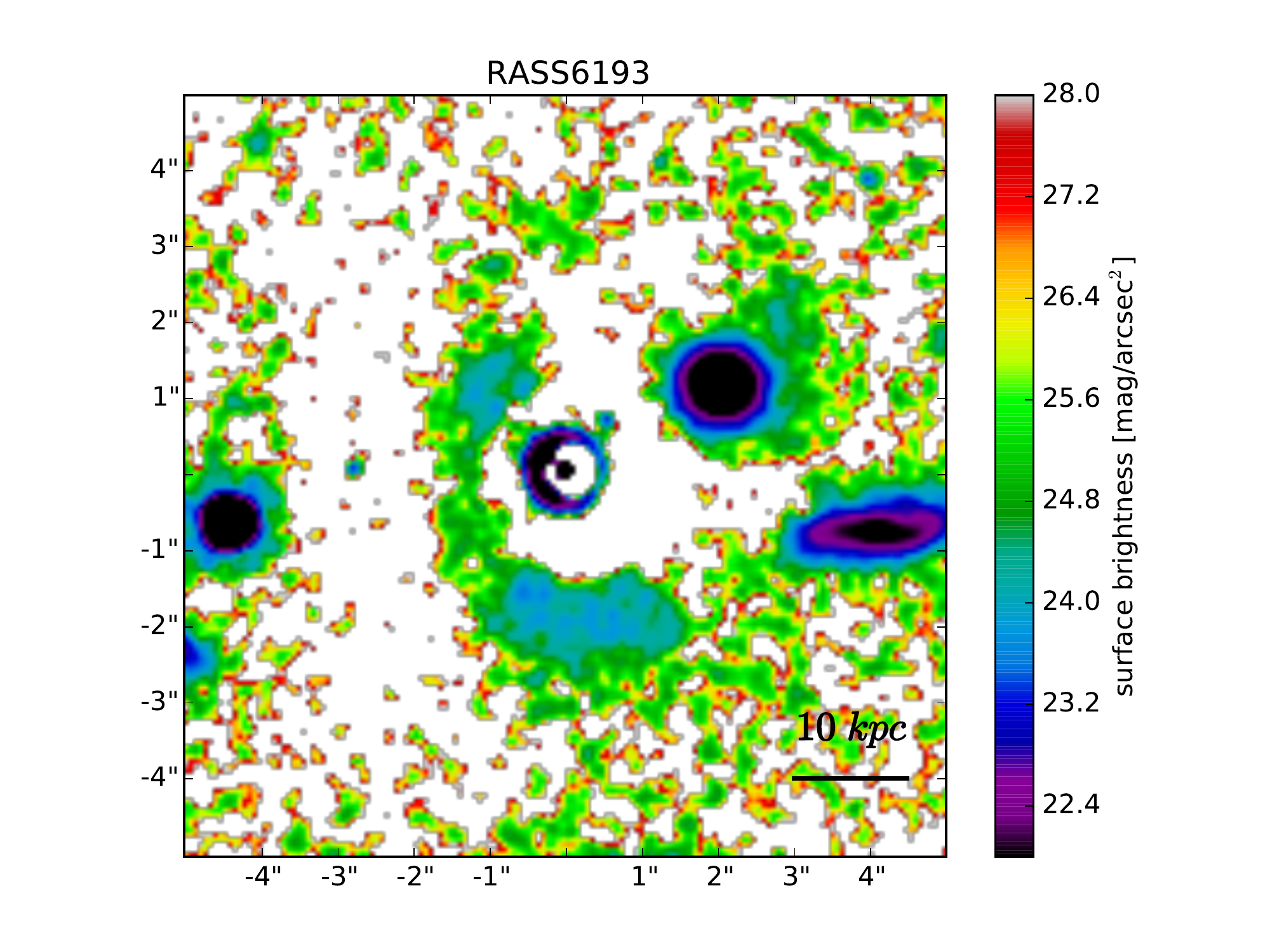}
\caption{Surface brightness maps showing residual extended merger features in RASS 1345, 2019, 4339, 6156 and 6193. Maps show surface brightness of features after subtraction of point source and galaxy flux as well as sky. Individual cut-outs are 10\arcsec x 10\arcsec and are oriented as observed rather than North up and East for consistency with other figures.}
\label{F:sb_merger}
\end{center}
\end{figure*}

The merger features that are detected show primarily as shells and fainter tidal tails with observed surface brightnesses $\sim 25\  \textrm{mag/arcsec}^{2}$ (see Fig. \ref{F:sb_merger}), with the most common feature being shell-like structures with extents of $\sim 10$kpc within $<50$kpc of the centre (seen in RASS4339, RASS6156, RASS6193). We compare the observed morphological features to simulations  by \citet{ji_lifetime_2014} for the surface brightnesses as well as \citet{barnes_transformations_1992} for qualitative features. Taking both k-corrections (assuming standard stellar populations) and surface brightness dimming into account, our data should be equivalent or deeper than the $\sim 28\  \textrm{mag/arcsec}^{2}$ optical simulations by \citet{ji_lifetime_2014}. This implies that our data is deep enough to detect major ($>$1/3) mergers up to $\sim 1$ Gyr after coalescence. We find that the the shell features seen in the majority of cases are qualitatively consistent with major mergers $< 1 Gyr$ after coalescence \citep[e.g.][]{ji_lifetime_2014}. Two sources (RASS 1345, RASS2019) show tidal features at similar surface brightnesses of about  $\sim 25\ \textrm{mag/arcsec}^{2}$, more consistent with features in simulations seen near coalescence. We detect no ongoing, highly disturbed mergers in our samples (i.e., such showing double-nuclei), however, given that this phase has a short duration compared to the merger time-scales, this is not entirely surprising, it does however argue against a large fraction of AGN being triggered (and visible as unobscured AGN) during coalescence or shortly thereafter.

Five sources in our sample do not have detected host galaxies, while they were included in the visual classification, no signs of mergers and disturbance were flagged in any of the unresolved sources. We therefore explore if the bright point sources in those cases might mask merger features. To do so, we add the residuals (i.e. tidal tails and shells left after galaxy subtractions) from the five sources classified as merging or disturbed above (see Fig. \ref{F:sb_merger}) to the residuals from the PSF fits for the unresolved sources and inspect the resulting image to determine detectability of typical merger features. We find that for the unresolved AGN RASS3845, the point source residuals were relatively weak, so all added merger features were recovered, the remaining sources showed strong PSF residuals, in those cases only the large, relatively high surface brightness tidal tail in RASS1345 was recovered. The bright AGN might therefore conceal later stage weaker merger features, such as the shells seen in our sample, as well as more narrow tidal tails, such as seen in RASS2019 (see Fig. \ref{F:sb_merger}). 

Applying these results to the data shown in Figure \ref{F:visclass}, if we assume that merger features (visual classification M) are hidden in $\geq 4/5$ unresolved AGN, the merger fraction in the AGN would be higher than that found in the control sample (at a 1$\sigma$ significance). However, as discussed above, this is a very extreme "best-case scenario" that would likely not be consistent with the simulation performed.  Similarly, if $\geq 3/5$ unresolved AGN would contain interaction features classified as disturbed, this would result in a \textit{marginally} statically significance difference (again, at 1$\sigma$ significance). However, all these scenarios would assume a drastically different morphology in the unresolved sample compared to the resolved sample, which is unlikely given the relatively minor difference in luminosities and black hole masses for the resolved and unresolved AGN. 

\begin{figure*}
\begin{center}
\includegraphics[width=8cm]{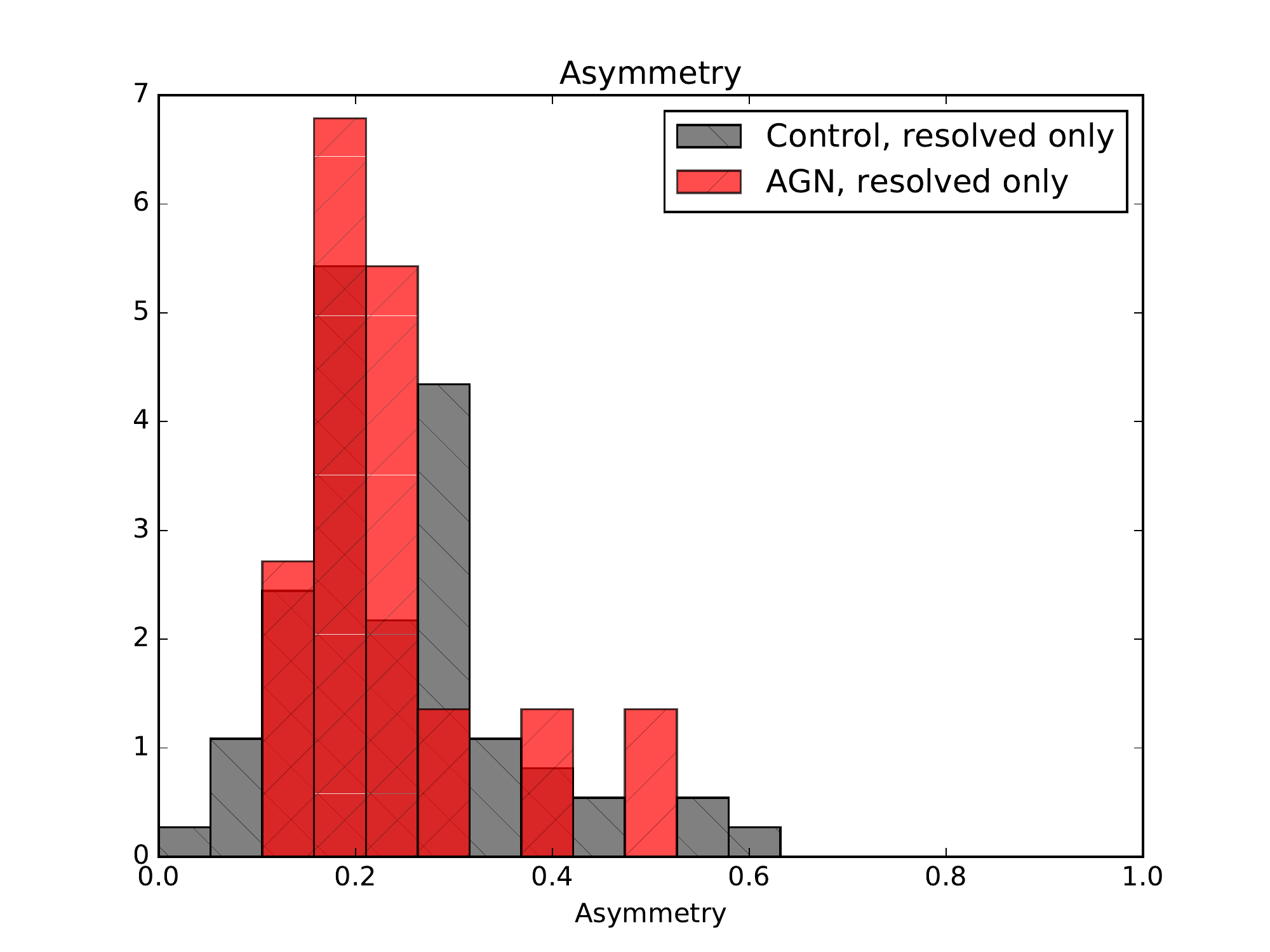}
\includegraphics[width=8cm]{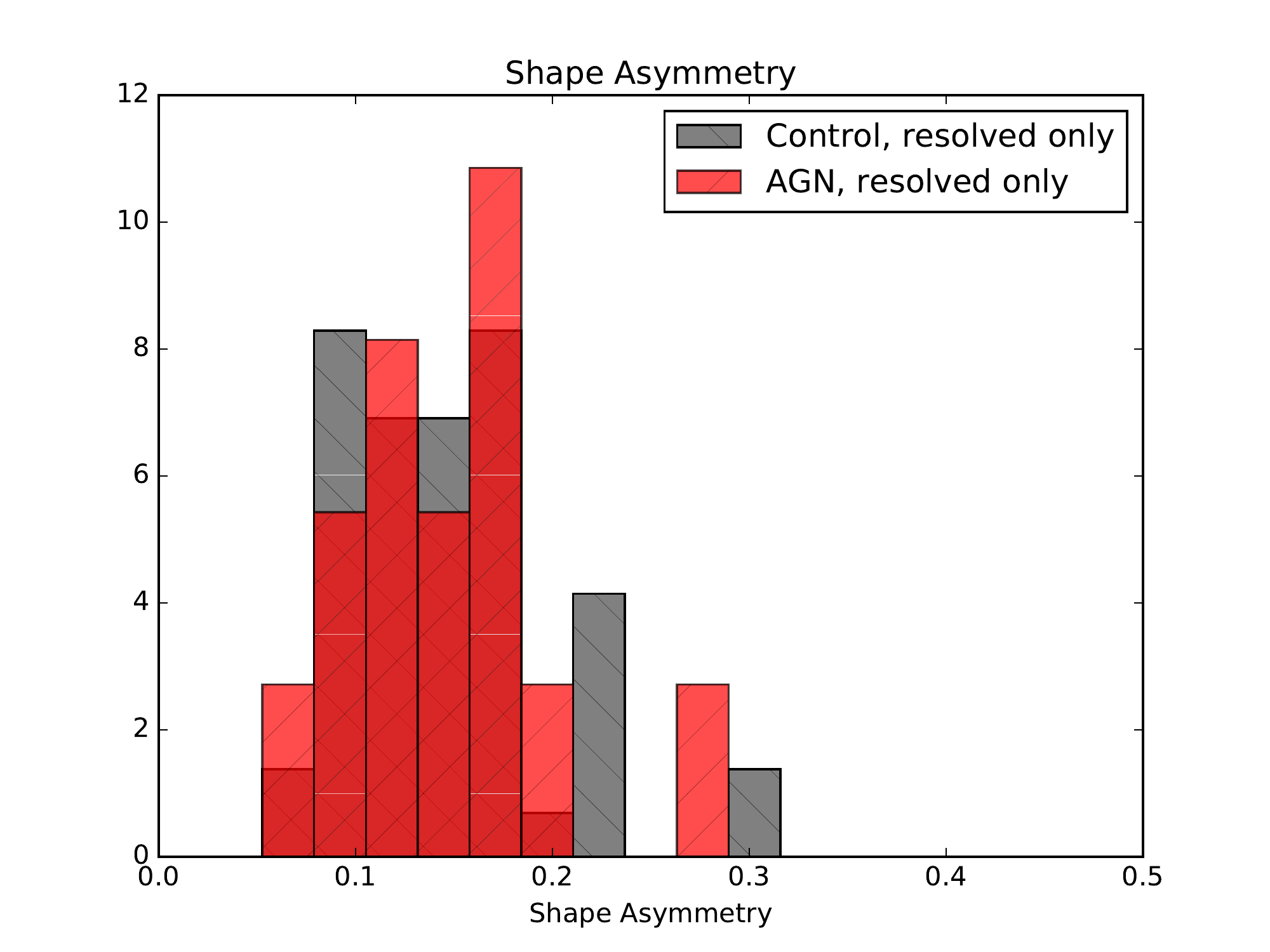}
\caption{Comparison for asymmetry (left) and shape asymmetry (right) of all resolved AGN hosts (red) and matched mock AGN control galaxies (grey).}
\label{F:a_hist}
\end{center}
\end{figure*}

Visual classification might miss more subtle differences between the samples. We therefore also compared the asymmetries of the two samples, see Figure \ref{F:a_hist} (left panel). There is no statistically significant difference between the two samples (p $>>$ 0.05 for both Mann-Whitney-U and KS tests). We do not find a higher rate of high-asymmetry galaxies in the AGN sample. Such a difference might not be picked up by the 2-sample test used since those are mainly sensitive to differences in the means of the samples.  Therefore, in agreement with previous work by this group on lower luminosity AGN \citep{villforth_morphologies_2014}, there is no suggestion of enhanced asymmetries in the AGN sample.

Similarly, the shape asymmetry, as introduced by \citet{pawlik_shape_2016} shows no statistically significant difference between the two samples (see Figure \ref{F:a_hist}, right panel, p $>>$ 0.05 for both Mann-Whitney-U and KS tests), as well as no higher fraction of sources with high asymmetry.

As for the visual classification, we would like to address the issue of unresolved AGN affecting the results. Assuming that all unresolved AGN have the highest asymmetry value found in both the AGN and control samples and recalculating the two-sample tests,  we find no statistically significant difference between the AGN and control sample for the standard asymmetry (p=0.1 for a Mann-Whitney U test\footnote{the Mann-Whitney U is used instead of the more common Kolmogorof-Smirnov test due to its greater sensitivity to changes in the tails of distributions}).

For the shape asymmetry, if we assume that all unresolved AGN have shape asymmetries as high as the highest value for the entire sample, we would find a statistically significant difference (p$<$0.005), however, in the more realistic case of assuming that the unresolved AGN are assigned the five highest values found in the entire sample, the difference is no longer statistically significant (p=0.11). It is therefore unlikely that the unresolved AGN hide a statistically significant difference between the AGN and control sample.

Both the visual classification and quantitative measures agree that there is no statically significant difference between the samples, suggesting that levels of disturbances in the host galaxies of luminous AGN are not enhanced, this finding extends previous work by this group on lower luminosity AGN \citep{villforth_morphologies_2014}. The limited efficiency of merger feature detections in the unresolved sample (5/20 sources) are unlikely to be responsible for the lack of differences observed between the AGN sample and control. Reasonable assumptions about the properties of those sources lead to no statistically significant difference between AGN and control sample.

\section{Discussion}
\label{S:discussion}

We performed 2D morphological fits to detect the host galaxies of 20 luminous AGN ($L_{bol} \sim 10^{46}$ erg/s) at redshift z$\sim$0.6. 15/20 host galaxies are resolved. We compare the host galaxy morphologies to a sample of control galaxies matched in absolute host galaxy magnitude at $\lambda_{rest} \sim 1\mu$m. We find no statistically significant difference between the morphologies of the AGN host galaxies and matched control, either using visual classification (see Figure \ref{F:visclass}) or quantitative morphology measures (see Fig. \ref{F:a_hist}). For the quantitative morphology measures, the AGN and control sample have almost identical mean values and the two-sample tests reveals no statistically significant differences. The morphological properties of the unresolved host galaxies would have to differ significantly from the resolved sources to be consistent with a higher merger fraction in the AGN sample. Therefore, in summary, we find no signs of enhanced morphological disturbances in the hosts of luminous AGN.

\subsection{Constraints on major merger rates}

While merger fractions are relatively poorly defined, we can assume an approximate major merger fraction from visual classifications of $< 20\%$ with no enhancement compared to control. We will now discuss detectability of merger features in this study.

A long time delay between the merger and onset of AGN activity would cause merger features to fade, making them non-detectable in relatively shallow studies beyond the local universe. Such time delays are supported by some studies showing a delay between starbursts and AGN activity of $\sim$ 250 Myr \citep[e.g.][]{wild_timing_2010}. We estimate that merger features in our sample should remain visible for $\sim$ 1 Gyr \citep{lotz_effect_2010,lotz_effect_2010-1,mihos_morphology_1995,elmegreen_smooth_2007,ji_lifetime_2014}, see also \citet{villforth_morphologies_2014}. However, merger mass ratios strongly influence the detectability, given the depth of our data, we would expect to detect major mergers up to a mass ratio of  $\mu_m \geqslant 0.3$ up to $\sim$ 1 Gyr after coalescence while minor mergers $\mu_m \geqslant 1/6$ will not be detectable post coalescence \citep{ji_lifetime_2014}. \citet{cibinel_physical_2015} analysed the the efficiency of quantitative merger tracers (in particular, A and $M_{20}$) in detecting merger features using a mixture of simulations and observational data, they generally found that Asymmetry has a high completeness (50-70\% at wavelengths comparable to those in our study) as well as low contamination (10\%). Since a comparison is made to a control sample, the contamination number does not greatly affect our results. The completeness estimated by \citet{cibinel_physical_2015} (50-70\%) indicates that any upper limit we set on merger enhancements, can be underestimated by a factor of 1.4-2 before completeness corrections are applied. \citet{cibinel_physical_2015} showed that Asymmetry as efficient at detecting mergers near coalescence, while the shape asymmetry developed by \citet{pawlik_shape_2016} traces later stage mergers.

In the result Section \ref{S:results}, we estimated that from the visual classification, we can place an upper limit on the enhanced merger fraction in the AGN sample compared to control of 11/27\%, for clear mergers and mild disturbances respectively, at 3$\sigma$ significance  (one-tailed p=0.05). This number needs to be adjusted for the completeness of the visual classification. We do not have exact numbers for these completenesses, however, given previous work on quantitative morphology measures \citep{pawlik_shape_2016}, it was shown that morphology measures can miss mergers identified by eye, we therefore expect the completeness to be higher than that derived by \citet{cibinel_physical_2015} (>50-70\%), leading to a correction of the upper limits by $<1.4$. From the visual classification, this leads to a very conservative upper limit of 15\% for clear mergers.

For the two quantitative measures, we can estimate upper limits on enhanced merger fractions in two ways. First, we can treat the quantitative measures as binary decisions, i.e., given the sample size of 20 for the AGN, how many AGN what is the upper limit of AGN in the high-asymmetry group. Due to the asymmetry of the binomial distribution, this depends on the cut-off used, but assuming a cut-off of 10\% for the general population, this gives the a 3$\sigma$ significance  (one-tailed p=0.05) upper limit of 16\% for the enhanced merger fraction. Taking into account the worst case completeness correction from \citep{cibinel_physical_2015} gives a conservative upper limit of 32\% on the enhanced merger fraction. This however discards significant information contained in the shape of the distribution and is therefore likely overestimated.

To take full advantage of the shape of the distributions, we simulate fake AGN populations asymmetry distributions. These fake samples consist of n values drawn from a normal distributions with values $\mu$ and $\sigma$ and 20-n values drawn from the asymmetry values for the control sample, yielding a sample size matching this study. These simulations therefore assume that a subpopulation of the AGN sample are drawn from a high-A distribution. Mann-Whitney U and KS tests are then run against the control sample. The Mann-Whitney test shows a lower fraction of false negatives, so we will discuss Mann-Whitney U tests only. We find that when the contaminant distribution is offset  by $\geqslant 1 \sigma$ from the control population, fake merger fraction $\geq10\%$ yield p-values $\leqslant0.1$. While the exact distribution of asymmetries in mergers for the given dataset is not known, we can therefore estimate an upper limit of $\sim 10\%$ from the quantitative data, taking completeness into account, this yields an enhanced merger fraction of $\leqslant 20\%$ in the AGN sample. While the exact offset of the enhanced asymmetry distribution affects the results, the influence of the size of the sub-sample is stronger. 

We therefore have upper limits on the contribution of recent major mergers on our sample, $<$15\%/38\% for clear mergers and mild disturbances from visual classification as well as $\leqslant20\%$ from quantitative morphology measures. 

The measures discussed above trace mostly recent major mergers. In the local universe, \citet{ramos_almeida_are_2011} find that very deep imaging of local radio galaxies reveals an enhancement of faint tidal tails and shells when compared to control. \citet{ramos_almeida_are_2011} show that these features are consistent with either a long time delay or minor mergers. Similarly, \citet{bennert_evidence_2008} showed that deeper imaging revealed merger features in galaxies previously believed to be undisturbed based on shallower imaging by \citet{dunlop_quasars_2003}. Unlike these lower redshift studies, we are unable to constrain long delays or minor mergers.

Additionally, the prevalence of disk galaxies in the sample speaks against a large fraction of host galaxies being remnants of wet major mergers, since those would likely result in spheroidal morphologies \citep[e.g.][]{barnes_transformations_1992,cox_kinematic_2006,hernquist_structure_1992,hernquist_structure_1993}.

\subsection{Comparison to theoretical predictions}

We will first discuss models in which major wet mergers are either the dominant triggering process for AGN \citep[e.g.][]{silk_quasars_1998,di_matteo_energy_2005,wyithe_physical_2002} or alternatively either dominate only at high luminosities \citep[i.e. high fuel masses, e.g.][]{somerville_semi-analytic_2008,hopkins_we_2013} or high black hole masses \citep[e.g.][]{hopkins_characteristic_2009}. Such merger-triggered AGN models are used in both semi-analytic \citep[e.g.][]{somerville_semi-analytic_2008}, hydrodynamical simulations \citep{hopkins_cosmological_2008} as well as purely theoretical calculations \citep{silk_quasars_1998} to reproduce the quasar luminosity function, clustering results as well as the abundance of massive quiescent galaxies at low redshift. Unlike previous work \citep[e.g.][]{boehm_agn_2012,gabor_active_2009,grogin_agn_2005,cisternas_bulk_2011,villforth_morphologies_2014,kocevski_candels:_2012}, the objects in this study all lie in the range that is assumed to be dominated by merger triggering in these theoretical models. As discussed above, our data do not show enhancements of merger features compared to control, indicating that major mergers are only responsible for a small fraction of even the most luminous AGN. The results from this study can be consolidated only if a very long time-delay between the merger and AGN phase is assumed. Such a delay would mean that AGN feedback cannot be responsible for the quenching of star formation following a major merger, as suggested in the theoretical models considered here \citep{sanders_ultraluminous_1988,di_matteo_energy_2005,hopkins_cosmological_2008,bower_breaking_2006,somerville_semi-analytic_2008}. Additionally, the prevalence of disk galaxies in the sample speaks against this scenario.

Related to the issue of long time delays washing out merger features is the suggestion that the selection of unobscured objects could bias this study towards late stages of a suggested quasar evolution phase in which the major merger leads to an "evolutionary sequence" starting with a heavily obscured quasar phase followed by a blow-out and appearance of an unobscured quasar \citep[e.g.][]{sanders_ultraluminous_1988,di_matteo_energy_2005,hopkins_cosmological_2008}. Some studies suggest that heavily reddened or X-ray obscured AGN show signs of enhanced merger features \citep{urrutia_evidence_2008,kocevski_are_2015,fan_most-luminous_2016}. However, if we assume that this obscured reddened quasar population evolves into the unobscured population, the lifetimes of the obscured phase would have to be long ($\sim$ 1 Gyr, see above) to accommodate the lack of merger features in this study. To agree with quasar lifetime estimates and the suggestion of an early triggering of AGN in the obscured phase, quasars would need to spend the majority of their lifetime in a heavily reddened or obscured phase. This would be both inconsistent with the high fraction of mergers posited in some obscured samples \citep{urrutia_evidence_2008} and the ratio between obscured and unobscured sources which also has to account for orientation dependent obscuration. An evolutionary sequence therefore does not naturally explain the lack of merger features in this and other unobscured AGN samples. On the other hand, in a series of papers, \citet{bussmann_hubble_2009,bussmann_hubble_2011} compared the morphologies of dust-obscured galaxies with and without strong AGN contribution at redshift z$\sim$ 2 and found significant differences, consistent either with a delayed onset of AGN activity after a merger-driven starburst or a different physical origin.  While delays might explain the differences, heavily obscured, merger associated AGN could form a sub-sample of the general AGN population without an evolutionary connection and rare enough to not show up in most AGN host galaxy studies. If obscured AGN are indeed triggered by major mergers, while unobscured AGN are not, the question of what processes trigger unobscured AGN remains.

A potential limitation to studies comparing host galaxies of AGN to control samples is fast AGN variability \citep[e.g.][]{gabor_simulations_2013,hickox_black_2014,king_agn_2015,schawinski_active_2015}. If AGN vary on short time-scales, the control sample could contain "recent AGN", limiting the information contained in comparisons to control samples. Even if we assume very short AGN lifetimes, the findings that major mergers are very rare in this sample (1/20) and the host galaxies are predominantly disk-like speaks against major mergers triggering the vast majority of AGN ($<$5\% considering clear major mergers). More quantitatively, we can calculate the detected merger fraction $r_{merger}$ assuming fast flickering.  We consider the fraction of galaxies experiencing a trigger $f_t$ as well as the duty cycle of AGN activity during a trigger phase $d_t$.  The fraction of mergers in the AGN sample is then:

\begin{equation}
r_{merger, AGN} = \dfrac{ d_{merger} f_{merger} }{ \sum d_t f_t}
\end{equation}

where $d_t / f_t$ refers to all possible triggers. Whereas the rate of mergers in the control sample is:

\begin{equation}
r_{merger, Control} = \dfrac{(1-d_{merger}) f_{merger}}{f_{no trigger} + \sum (1-d_t) f_t}
\end{equation}

A theoretical model that assumes that mergers are the dominant trigger for an AGN population is one where $d_{merger} >> \sum d_t$.  An AGN dominant model therefore results in a large detected merger fraction in the AGN sample and low detected merger fraction in the control sample. Flickering has little effect on this result. The only case in which flickering results in a comparable detected merger fraction in the AGN and control sample in which the duty cycle of AGN in mergers is either low $d_{merger} \leqslant \sum d_t$ or the overall fraction of mergers is low compared to other triggers $f_{merger} << \sum f_t$. However, in these cases. mergers have an insignificant effect on the triggering of AGN. Treating flickering as a duty cycle therefore shows that short life times to not significantly influence our results unless the overall contribution of major mergers is low.

Our results are therefore inconsistent with a view in which major mergers are the dominant triggering mechanism at high AGN luminosities ($L_{bol} > 10^{45} erg/s$), even taking flickering and completeness limitations into account.

Additionally, we can consider models convolving dark matter merger rates and AGN light curves and allowing a range of merger rates to trigger AGN \citep[e.g.][]{wyithe_self-regulated_2003,lapi_quasar_2006,shen_supermassive_2009,shankar_constraints_2010}. At any given epoch the seed black hole in each dark matter halo is allowed to grow at Eddington or even super-Eddington rates until a peak luminosity set by a quasar-feedback condition determined by the potential well of the host dark matter halo \citep[e.g.][]{granato_physical_2004}. The post-peak luminosity is assumed to decrease as a power law mirroring the gradual decrease in the available gas in the host galaxy. The observed AGN luminosity function at a given epoch $z_{sh}$ and $L_{bol}$ is thus contributed by all merger events triggered at $z_{sh}$ and shining with $L=L_{bol}$ at $z_{sh}$. \citet{shen_supermassive_2009} showed that dark matter halo merger rates above a mass ratio of $\mu_m=0.25$ could account for the whole AGN bolometric luminosity function at $z \geq 1$, and for the $L>L*$ AGN at $z<1$. These models predicts an equal weight of major and minor mergers at all luminosities, despite the input light curves and implied Eddington ratio distribution at fixed halo mass being very broad. While the specific model fit in \citet{shen_supermassive_2009} is consistent with our data following the line of argument above, lowering the mass ratio at which mergers trigger AGN would result in a distribution of merger rates triggering AGN skewed towards minor mergers, given the distribution of merger mass ratios of both halos and galaxies is found to be a powerlaw \citep{fakhouri_merger_2010,rodriguez-gomez_merger_2015}, mergers with mass ratios 0.1 outnumber mergers with mass ratios $>$0.25 by almost an order of magnitude \citep{rodriguez-gomez_merger_2015}. If we assume that AGN are triggered by mergers with mass ratios of 0.1 or below, major mergers that would be clearly visible in most datasets would form such a small fraction of the triggers, that no clear enhanced merger features would be expected.

Next, we will consider semi-empirical models based primarily on observational results. Based on AGN variability models and observed merger rates in starburst galaxies of different luminosities, \citep{hickox_black_2014} predict merger fraction of $\sim 40\%$ in the redshift and luminosity range of our sample. Note that this does not refer to the number of AGN triggered by mergers, but rather those found to have merger host galaxies while in an active state. At a merger rate $\leqslant$ 20 \%, our data are inconsistent with this result. \citet{draper_merger-triggered_2012} modelled AGN triggering mechanisms based on the number densities of massive gas-rich galaxies and predicted that merger triggering is only dominant at high redshifts and luminosities. Based on \citet{draper_merger-triggered_2012}, we would not expect the objects in our sample to be triggered by major mergers, since triggering by major mergers in this model starts to dominate only at z$\geqslant$1. At the redshift of our sample, \citet{draper_merger-triggered_2012} predict about 10\% of AGN to be triggered by major mergers, such a low rate is consistent with our data. 

Our results are therefore inconsistent with major mergers dominating AGN triggering at high AGN luminosities, both from the incidence of merger features and the high prevalence of disk-like morphologies in the sample. Our findings are consistent with a range of mass ratio minor mergers, as well as other secular processes triggering AGN.

\subsection{Comparison to other observations}

Our findings agree with a number of previous studies finding no enhancement of the incidence of merger features in samples of AGN over a wide range of redshifts, but at lower luminosities  \citep[e.g.][]{boehm_agn_2012,gabor_active_2009,grogin_agn_2005,cisternas_bulk_2011,villforth_morphologies_2014,kocevski_candels:_2012}.  Our results therefore show that the previous findings of a lack of enhanced merger features extends to very high AGN luminosities. Our study therefore also agrees with findings from \citet{mechtley_most_2015}, who studied the host galaxies of luminous AGN at z$\sim$2 and found no signs of enhanced merger features in the host galaxies of AGN.

The results from this study disagree with lower redshift studies \citep[][with redshifts $0.05$z$<0.7$, z$<<$0.1 and z$\sim$0.1 respectively]{ramos_almeida_are_2011,cotini_merger_2013,hong_correlation_2015}. All these studies analysed hosts of low redshift AGN selected either in the radio \citep{ramos_almeida_are_2011}, hard-Xray \citep{cotini_merger_2013} or optical \citep{hong_correlation_2015} and found increased incidence of merger features in the hosts of AGN compared to control. This could either be due to deeper imaging, which has shown to reveal merger features in AGN host galaxies not previously seen to be associated with mergers \citep{bennert_evidence_2008,smirnova_seyfert_2010,ramos_almeida_are_2011} or a different triggering mechanism as a function of redshift. The merger features in this sample have surface brightnesses down to $\mu_{SB} \sim 25.5$ mag/arcsec$^{2}$, see Fig. \ref{F:sb_merger}. \citep{hong_correlation_2015} are able to detect extended emission down to a surface brightness limit of $\mu_{SB} \sim 26$ mag/arcsec$^{2}$ at a comparable rest-frame wavelength, although at much lower resolution. \citep{cotini_merger_2013} use SDSS data which has lower surface brightness limit and resolution than our data. The discrepancy between the detection of merger features is therefore unlikely due to a difference in surface brightness limits. Another explanation for the discrepancy between low and high-redshift studies is a difference in triggering mechanisms: at low redshift, gas fractions in galaxies are generally low, so mergers might be the only viable mechanisms to produce large gas inflows towards galaxy centres, while at higher redshift, gas fractions in galaxies rise, leading to disk instabilities, which could fuel AGN.  While z=0.6 galaxies have considerably lower gas fractions than z=2 galaxies, their gas fractions are several factors higher than those in the local universe \citep[e.g.][]{geach_evolution_2011,genzel_combined_2015}. Alternatively, the radio selection used in \citet{ramos_almeida_are_2011} might favour merger host galaxies, in agreement with findings of high incidences of mergers in high-redshift radio galaxies by \citet{chiaberge_radio_2015}.

\subsection{What triggers AGN?}

We find a lack of an apparent enhancement of merger features in the hosts of luminous AGN beyond the local universe, in agreement with a range of other studies at lower luminosities  \citep[e.g.][]{boehm_agn_2012,gabor_active_2009,grogin_agn_2005,cisternas_bulk_2011,villforth_morphologies_2014,kocevski_candels:_2012,mechtley_most_2015}.  We argue that time delays are unlikely to explain the results, rather, AGN are likely triggered by a range of processes, including minor mergers. We argue that major wet mergers do increase the incidence of AGN, as for example found when studying AGN fractions in close galaxy pairs \citep[e.g.][]{ellison_galaxy_2011,ellison_galaxy_2013,koss_merging_2010}. However, since major mergers are rare, this limited increase in AGN fractions means that major mergers do not contribute significantly to the overall triggering rate. Major mergers trigger AGN, but AGN are not predominantly connected to major mergers. Rather, a range of secular processes as well as minor mergers could be responsible for triggering AGN. This is in agreement with local studies finding that a large fraction of mass and black hole growth is driven by minor mergers \citep{kaviraj_importance_2014}.

We also discussed studies of sub-samples of reddened \citep{urrutia_evidence_2008,kocevski_are_2015,fan_most-luminous_2016} and radio-loud AGN \citep{chiaberge_radio_2015} that show a high fraction of mergers.  This might suggest differences in AGN properties depending on triggering mechanisms, causing differences in host galaxy properties with AGN type. Additionally, large amounts of dust in starburst galaxies, might lead to higher obscuration in merger-triggered starbursts, causing higher merger rates in heavily obscured AGN. These would imply that obscured AGN connected to mergers are obscured on galaxy scale, rather than having different nuclear obscuration.

An interesting finding in this context is \citep{sabater_triggering_2015} who found that while interactions increase the fraction of AGN, this increase can be explained purely by the increase of the central gas density traced by the central star formation rate. \citep{sabater_triggering_2015} argue that mergers increase the central gas density, and that an enhanced central gas density increases the probability for a galaxy to host an AGN. This suggests that mergers provide a non-unique way to provide the necessary gas supply to the central kpc, but that once the gas is supplied by any process, mergers do not trigger the AGN. 

In summary, major mergers are not found to dominate AGN triggering at the highest AGN luminosities. Our data are consistent with triggering by less extreme processes such as minor mergers or secular processes.

\section{Conclusions}
\label{S:conclusions}

In this paper, we analysed the morphologies of the host galaxies of 20 luminous AGN ($L_{bol} \sim 10^{46}$ erg/s) and compared them to the morphologies of a control samples of galaxies matched in absolute $H/F160W$ band magnitude as well as redshift. Our findings can be summarized as follows:

\begin{itemize}
\item 15/20 AGN show clearly resolved host galaxies with magnitude differences $-1 \leqslant m_{gal} - m{qso} \leqslant 2$. The host galaxies have absolute magnitudes of $\sim -23.5$ in observed H band, effective radii $\sim 2.5-20$kpc with typical values of $\sim 5$ kpc and typical S\'{e}rsic indexes $1-4$, with about 2/3 of sources showing disk-like morphologies and 1/3 bulge-like morphologies. Host galaxies were matched to control galaxies in absolute magnitude and redshift and mock AGN were created by adding point sources matched in magnitude.
\item 4/20 sources show clear signs of disturbance, one source is a clear merger with extended tidal features, while three other sources show signs of disturbance at a lower level. Compared to the control sample, visual classifiers find no enhanced disturbances of merger features in the AGN sample when compared to control.
\item Both asymmetry \citep{conselice_symmetry_1997} and shape asymmetry \citep{pawlik_shape_2016} were measured for the resolved AGN and control sample. The AGN hosts do not show any signs of enhanced asymmetries, both in the traditional asymmetry \citep{conselice_symmetry_1997} that traces the bulk of the emission and the shape asymmetry \citep{pawlik_shape_2016} that is more sensitive to faint extended features such as tidal tails. 
\item The AGN with unresolved host galaxies are unlikely to hide a merger-AGN connection, we would have to assume radically different morphological properties between the resolved and unresolved AGN host galaxies in order to accommodate an enhanced merger rate in the AGN sample.
\item Taking into account the incompleteness of the morphology measures \citep{cibinel_physical_2015} as well we sample sizes, we estimate a conservative upper limit for enhanced merger fraction of $\leqslant 20\%$ from the quantitative analysis, and for the visual classification $<15\%$ for major mergers and $<38\%$ for mild disturbances.
\item In summary, we find no signs of enhanced interactions in a sample of luminous ($L_{bol} \geqslant 10^{45}$ erg/s) AGN at z $\sim$ 0.6. Theoretical models predict that at such luminosities, merger triggering should become dominant \citep[e.g.][]{somerville_semi-analytic_2008,hopkins_we_2013}. We find no signs that this is the case. The lack of merger features in the sample could still be consistent with either significant delays between the merger and onset of AGN activity of $\geqslant 1$ Gyr after coalescence or a prevalence of minor mergers. We argue that a long time-delay is unlikely to explain the results, instead, other triggering mechanisms dominate AGN activity up to the highest luminosities. This result is still consistent with major mergers being relatively efficient at triggering AGN, but too rare to significantly contribute to the AGN population.
\end{itemize}

The emerging picture that unobscured AGN, even at the highest luminosities, show no enhancement in merger signatures when compared to control \citep{boehm_agn_2012,cisternas_bulk_2011,grogin_agn_2005,mechtley_most_2015,kocevski_candels:_2012,villforth_morphologies_2014} has been expanded to the highest luminosities in this study. Host galaxies are predominantly disk-like, suggesting that they are not remants of major wet mergers. These result pose serious problems for simple models in which AGN appear at the end stages of mergers and shut down remaining star formation \citep[e.g.][]{sanders_ultraluminous_1988,di_matteo_energy_2005,hopkins_cosmological_2008}. Either other processes dominate AGN fuelling over large luminosity ranges, or a significant time delay exists between the major merger and onset of AGN activity. While a time delay would account for the lack of merger features observed, this poses serious problems for models requiring AGN feedback to shut down star formation \citep{bower_breaking_2006,somerville_semi-analytic_2008,croton_simple_2009,hopkins_cosmological_2008,silk_quasars_1998,di_matteo_energy_2005} as well as the prevalence of disk galaxies in the sample. Our results are consistent in a picture where minor mergers as well as secular processes trigger the majority of AGN at all luminosities.

\section*{Acknowledgments}
We thank the referee Frederic Bournaud for helpful and constructive comments that have improved the manuscript. FS warmly thanks Yue Shen for a number of valuable discussions, and for cross-checking some key predictions of merger rate models discussed in this work. Support for program number HST-GO-13305 was provided by NASA through a grant from the Space Telescope Science Institute, which is operated by the Association of Universities for Research in Astronomy, Inc., under NASA contract NAS5-26555. K.~R. acknowledges support from the
European Research Council Starting Grant SEDmorph (P.I. V.~Wild). M.M.P acknowledges support from the European Career Reintegration Grant Phiz-Ev (P.I V. Wild).'

\bibliographystyle{mn2e}
\bibliography{cycle21}

\newpage

\appendix

\begin{figure*}
\begin{center}
\includegraphics[width=18cm]{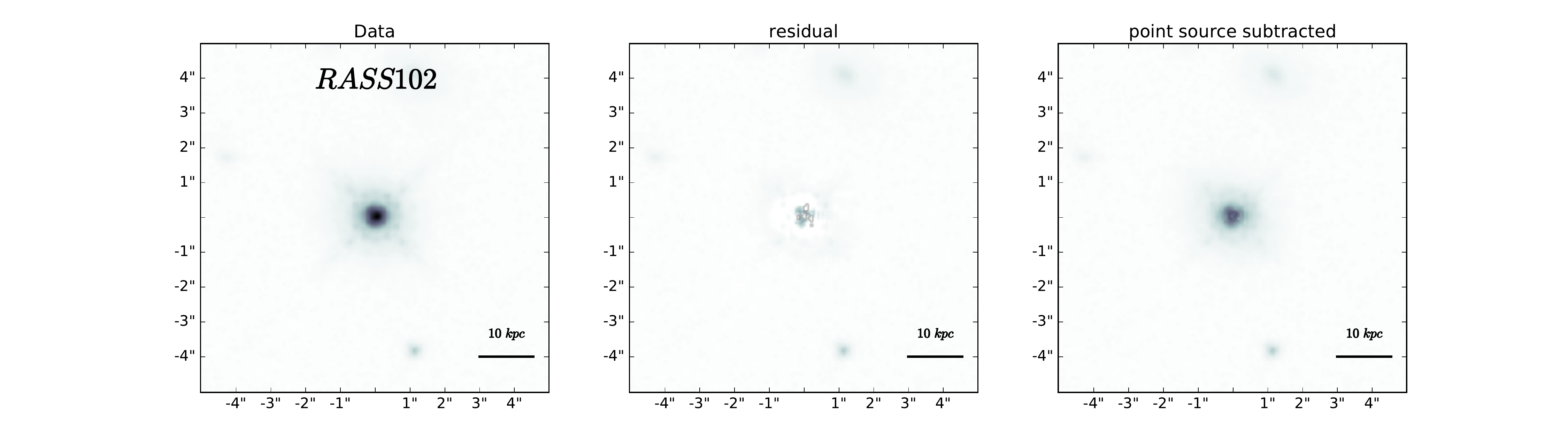}
\includegraphics[width=18cm]{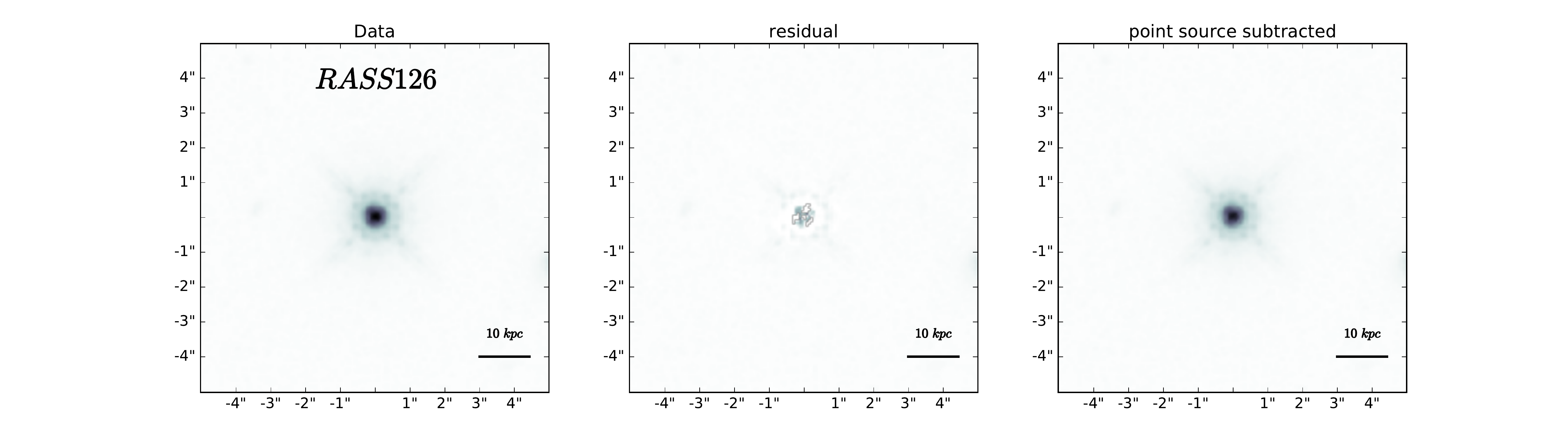}
\includegraphics[width=18cm]{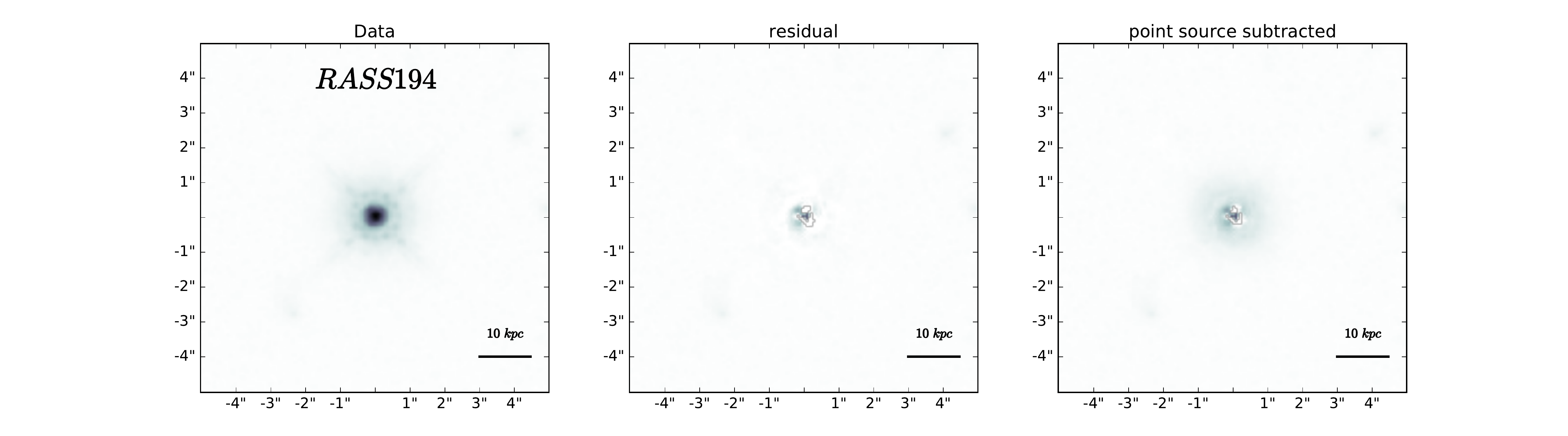}
\includegraphics[width=18cm]{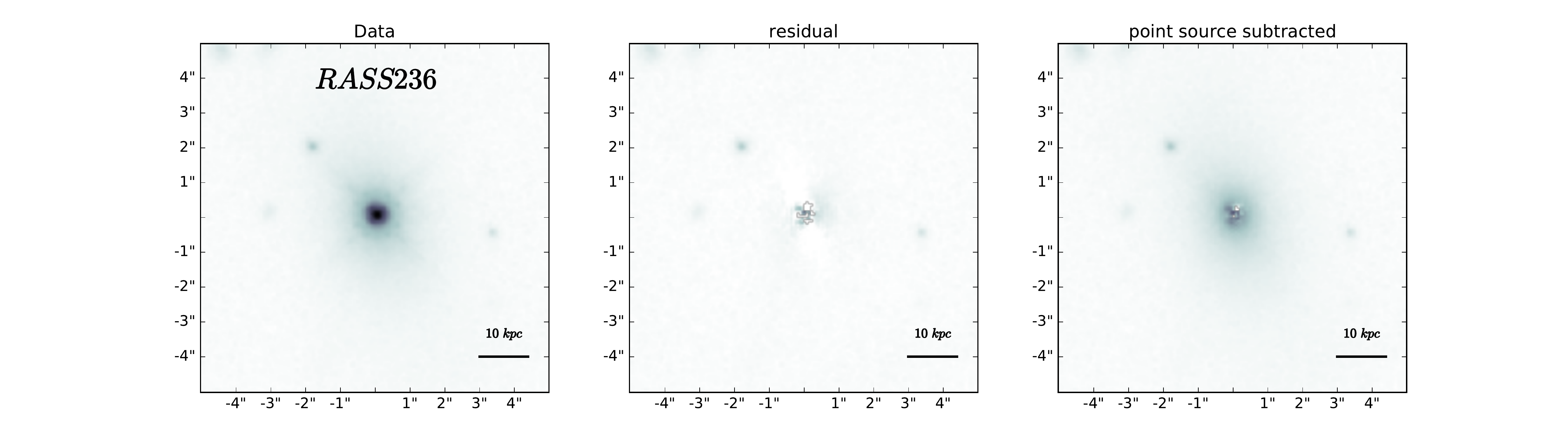}
\caption{\textsc{galfit} host galaxy fits for AGN in this sample, left most column shows raw images, middle column shows residual with full model subtracted, right most column shows residual with point source subtracted. Individual cut-outs are 10\arcsec x 10\arcsec and are oriented as observed rather than North up and East left to show the similarity in PSF patterns. }
\label{F:resid_1}
\end{center}
\end{figure*}

\begin{figure*}
\begin{center}
\includegraphics[width=18cm]{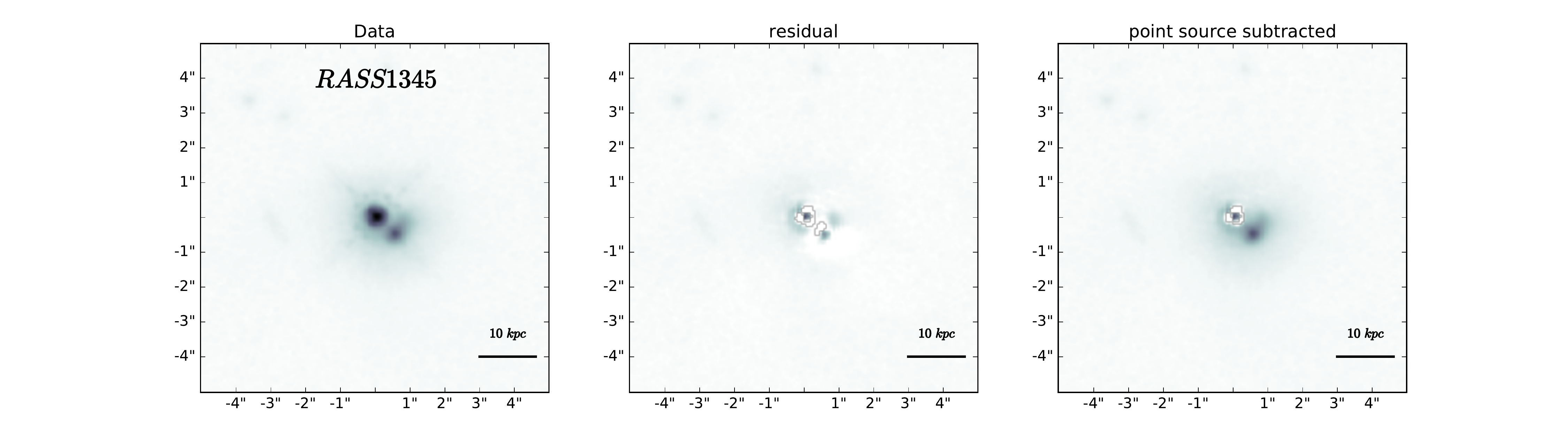}
\includegraphics[width=18cm]{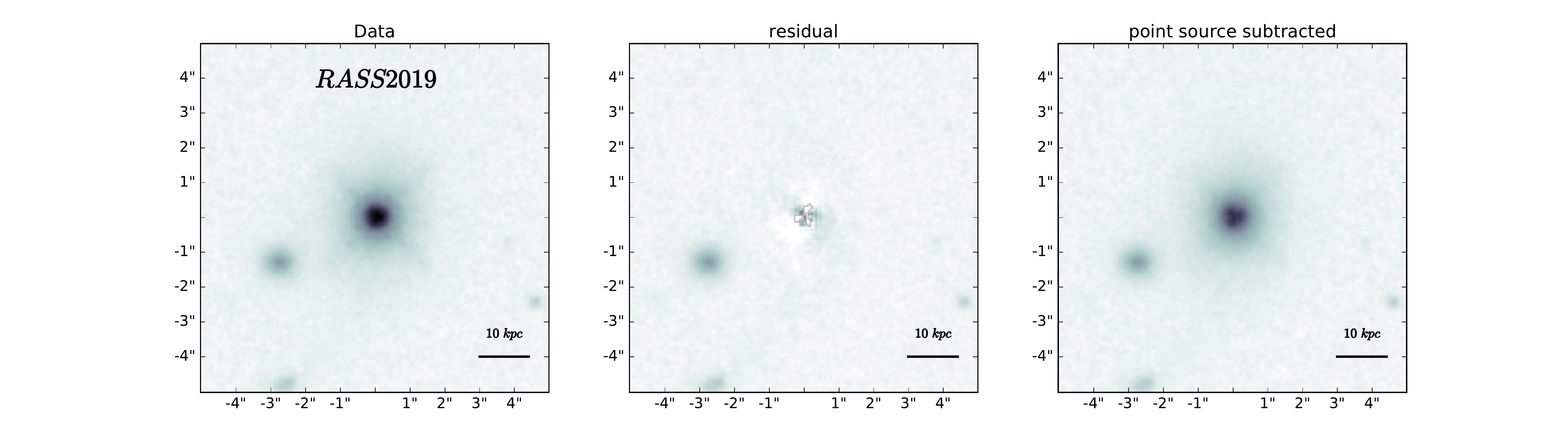}
\includegraphics[width=18cm]{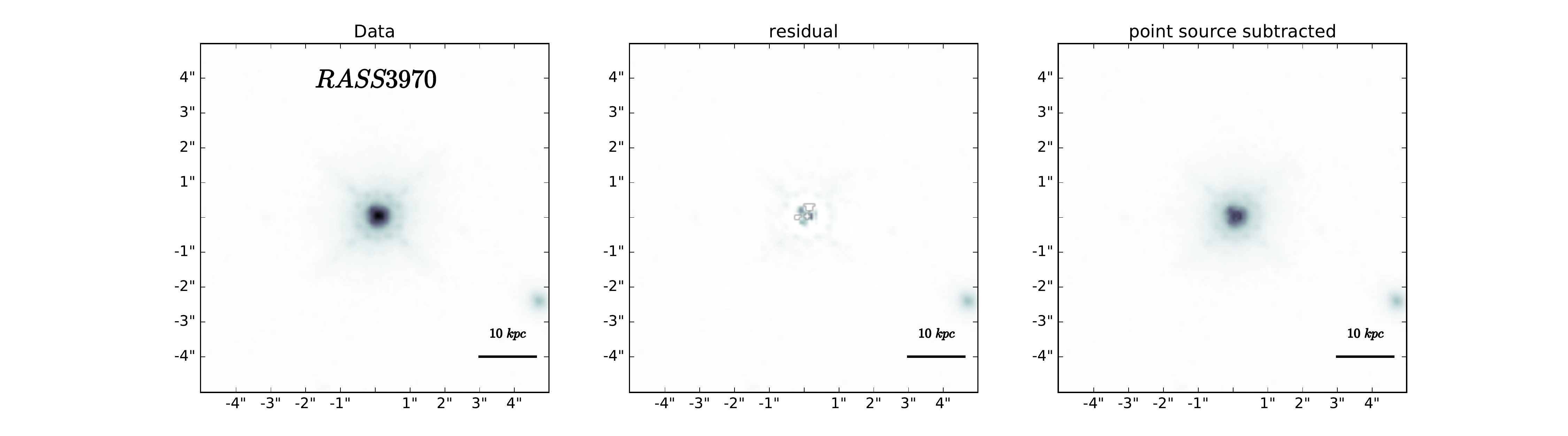}
\includegraphics[width=18cm]{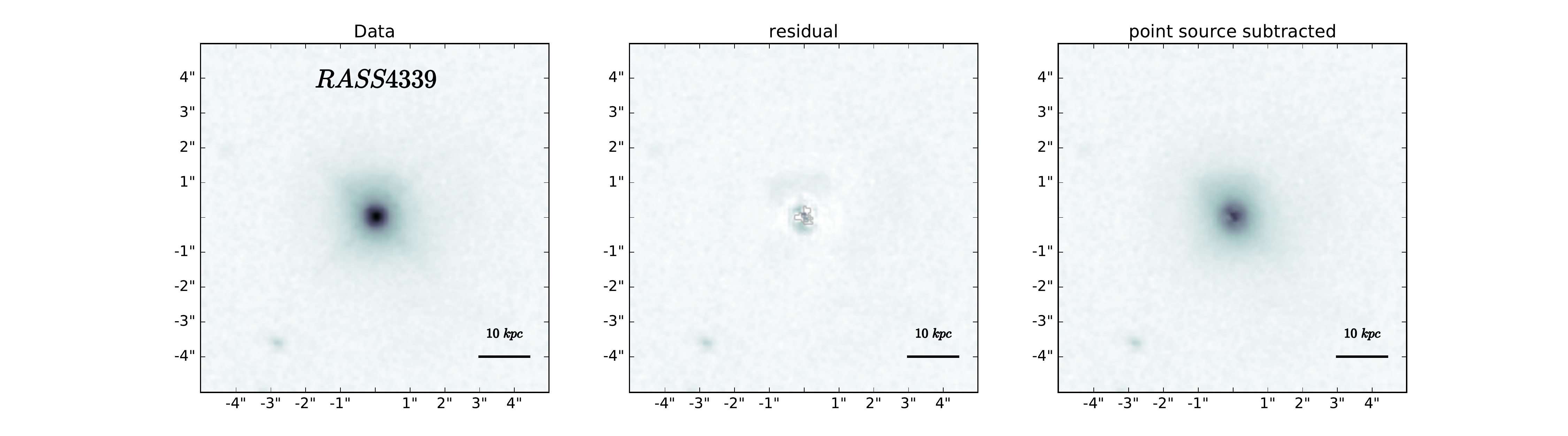}
\caption{\textsc{galfit} host galaxy fits for AGN in this sample, left most column shows raw images, middle column shows residual with full model subtracted, right most column shows residual with point source subtracted. Individual cut-outs are 10\arcsec x 10\arcsec and are oriented as observed rather than North up and East left to show the similarity in PSF patterns.}
\label{F:resid_2}
\end{center}
\end{figure*}

\begin{figure*}
\begin{center}
\includegraphics[width=18cm]{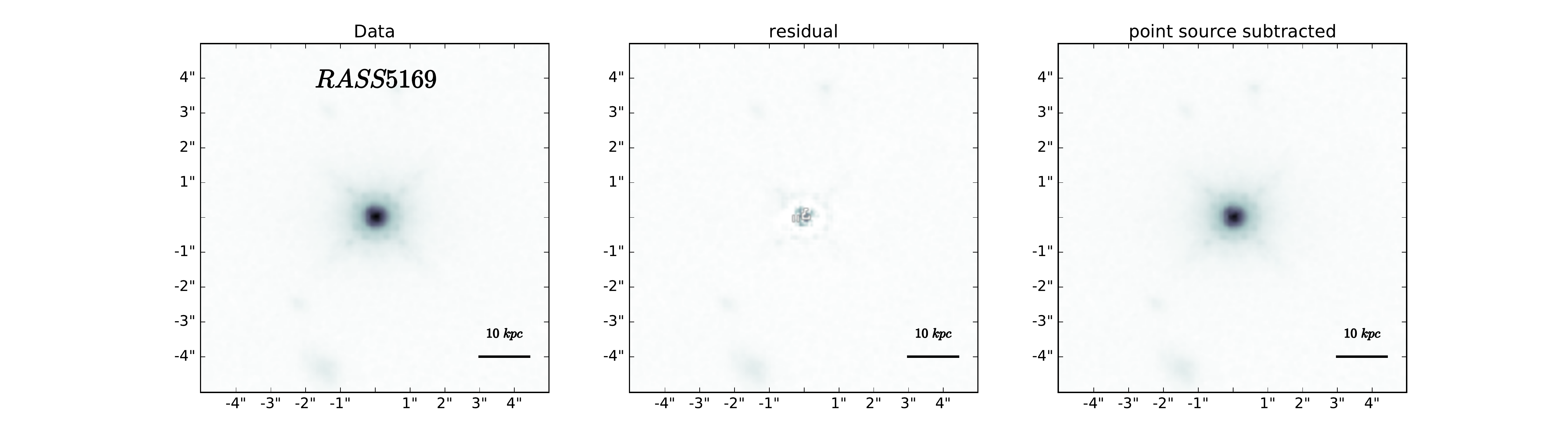}
\includegraphics[width=18cm]{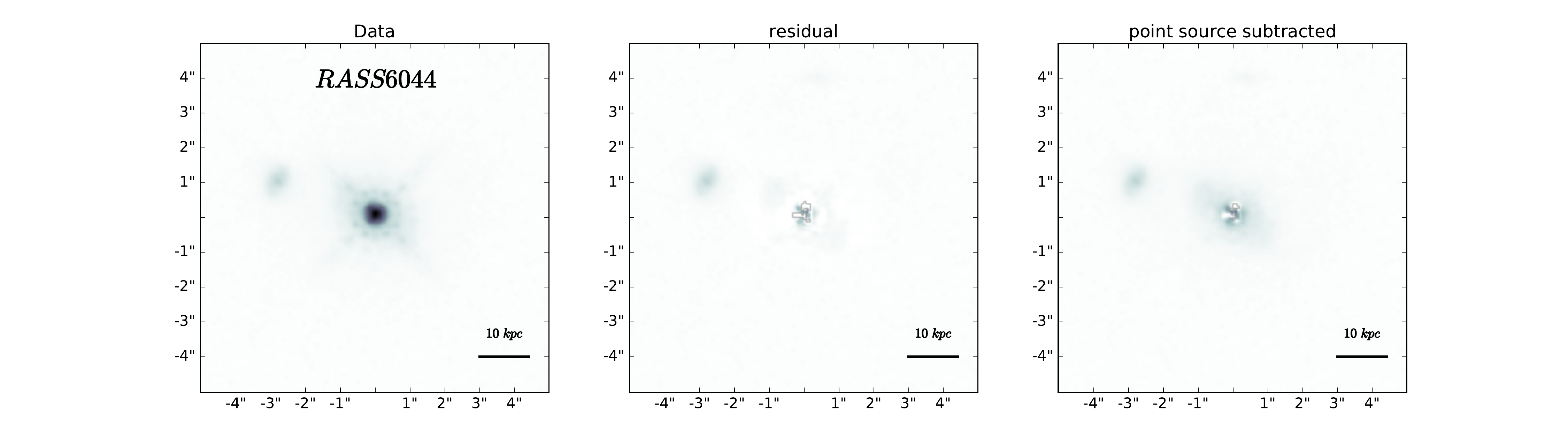}
\includegraphics[width=18cm]{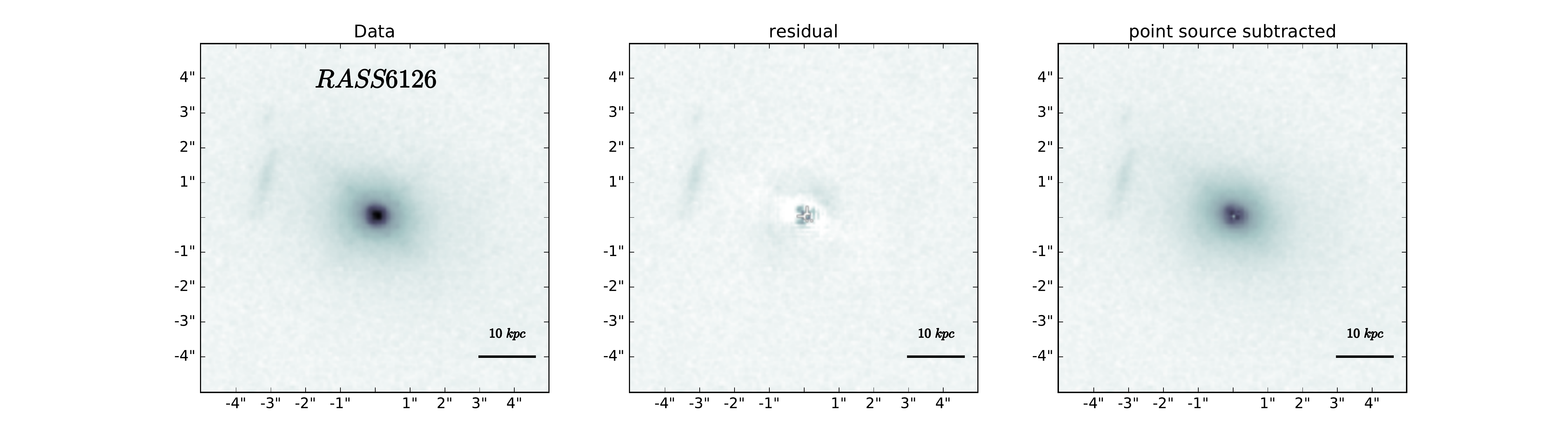}
\includegraphics[width=18cm]{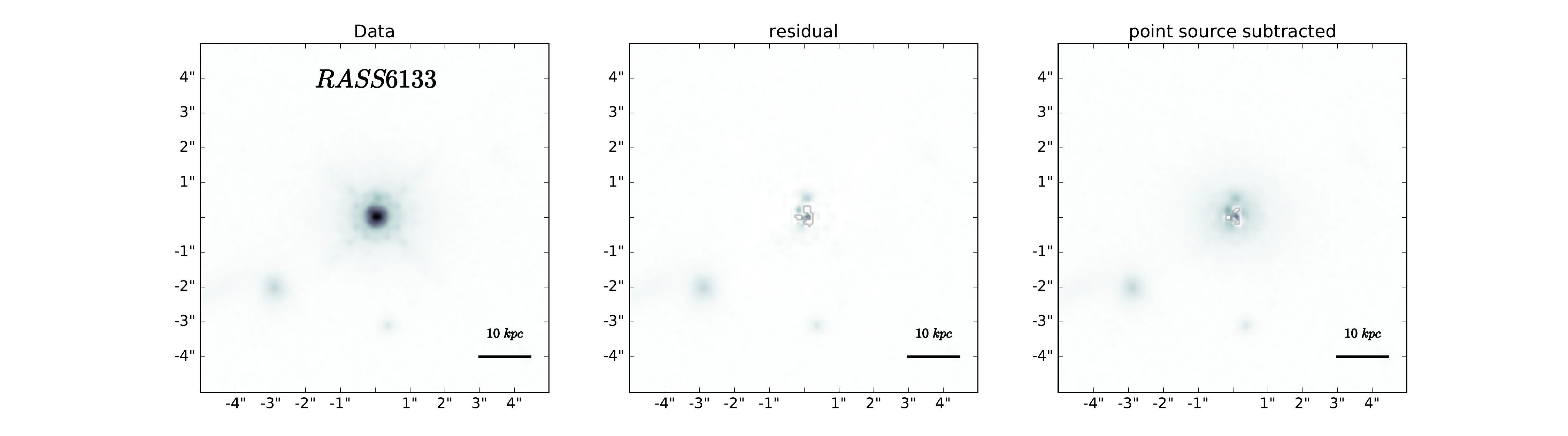}
\caption{\textsc{galfit} host galaxy fits for AGN in this sample, left most column shows raw images, middle column shows residual with full model subtracted, right most column shows residual with point source subtracted. Individual cut-outs are 10\arcsec x 10\arcsec and are oriented as observed rather than North up and East left to show the similarity in PSF patterns.}
\label{F:resid_3}
\end{center}
\end{figure*}

\begin{figure*}
\begin{center}
\includegraphics[width=18cm]{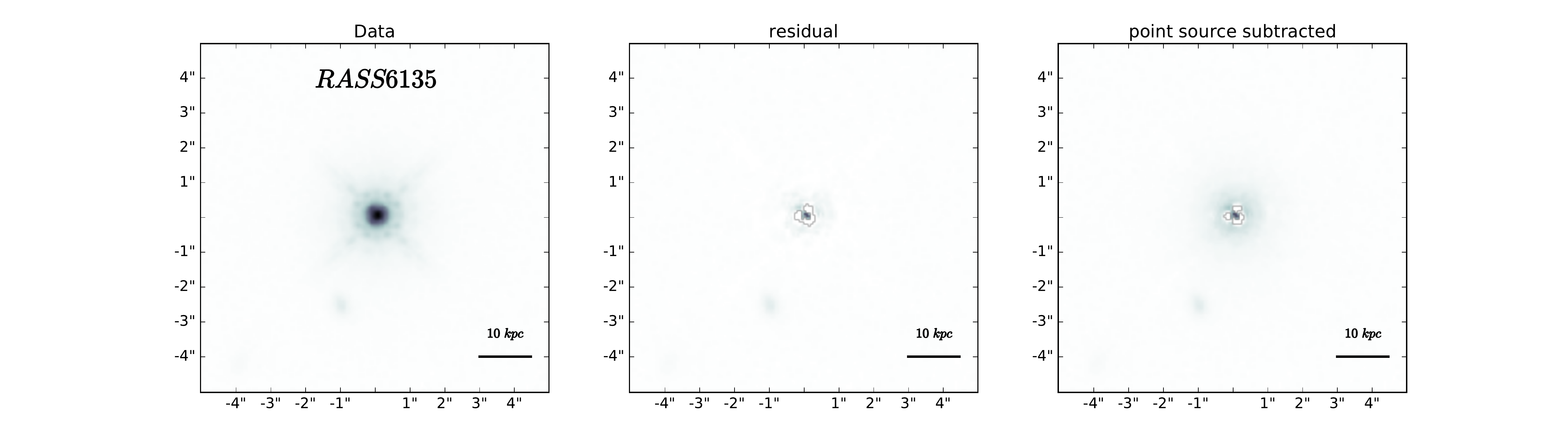}
\includegraphics[width=18cm]{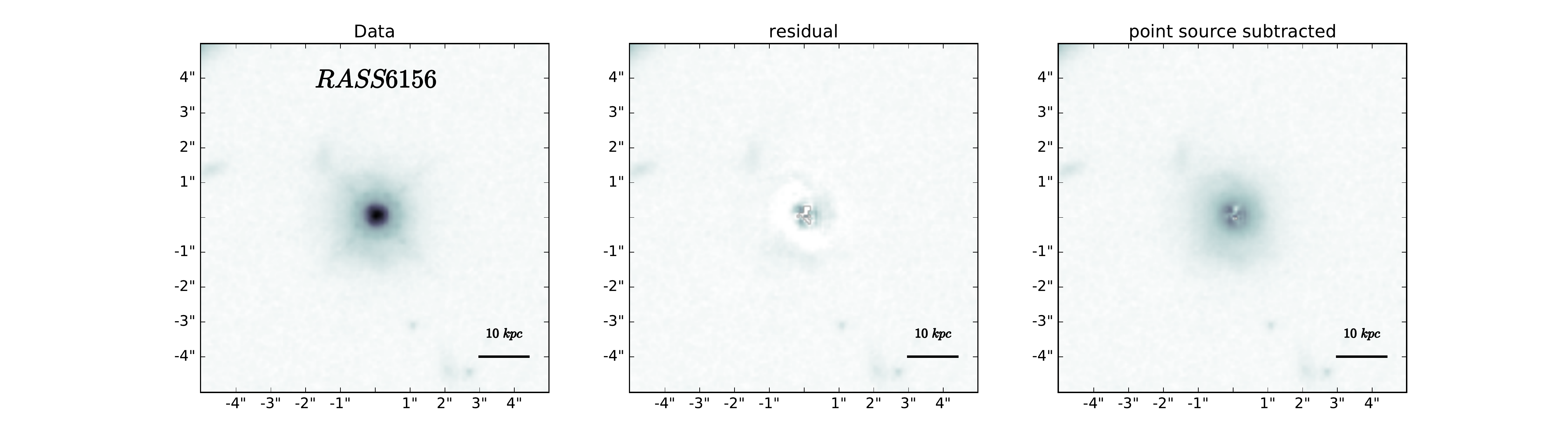}
\includegraphics[width=18cm]{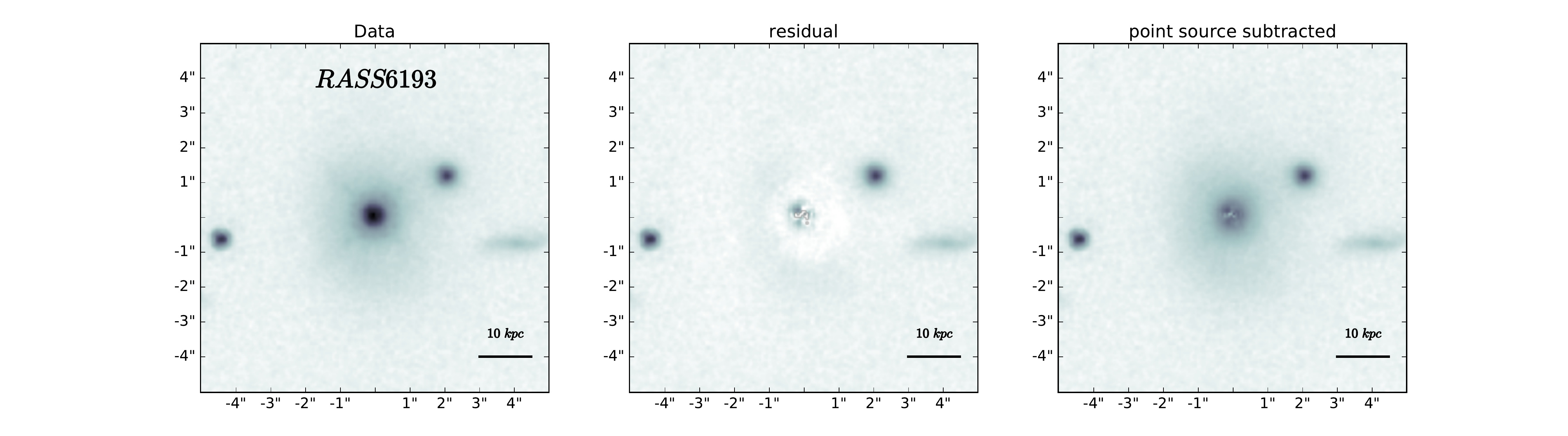}
\caption{\textsc{galfit} host galaxy fits for AGN in this sample, left most column shows raw images, middle column shows residual with full model subtracted, right most column shows residual with point source subtracted. Individual cut-outs are 10\arcsec x 10\arcsec and are oriented as oserved rather than North up and East left to show the similarity in PSF patterns.}
\label{F:resid_4}
\end{center}
\end{figure*}

\begin{figure*}
\begin{center}
\includegraphics[width=14cm]{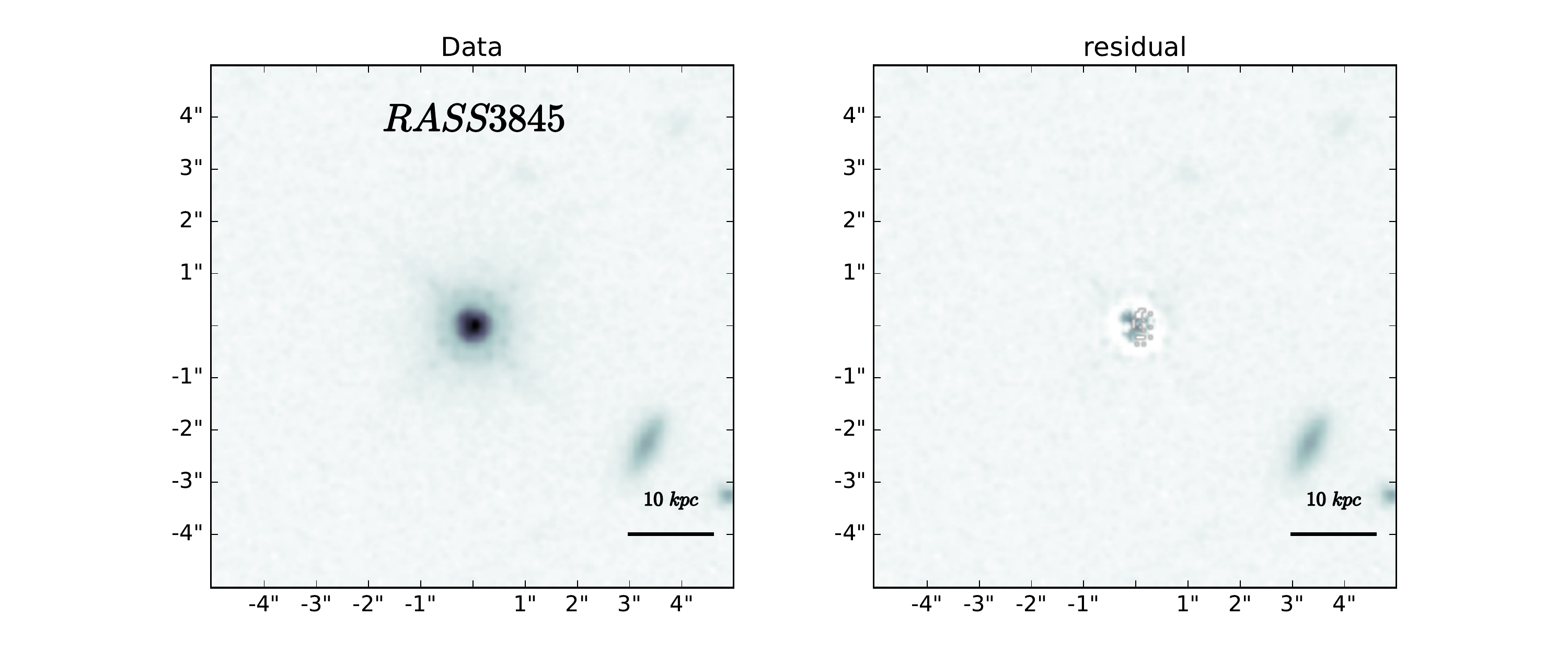}
\includegraphics[width=14cm]{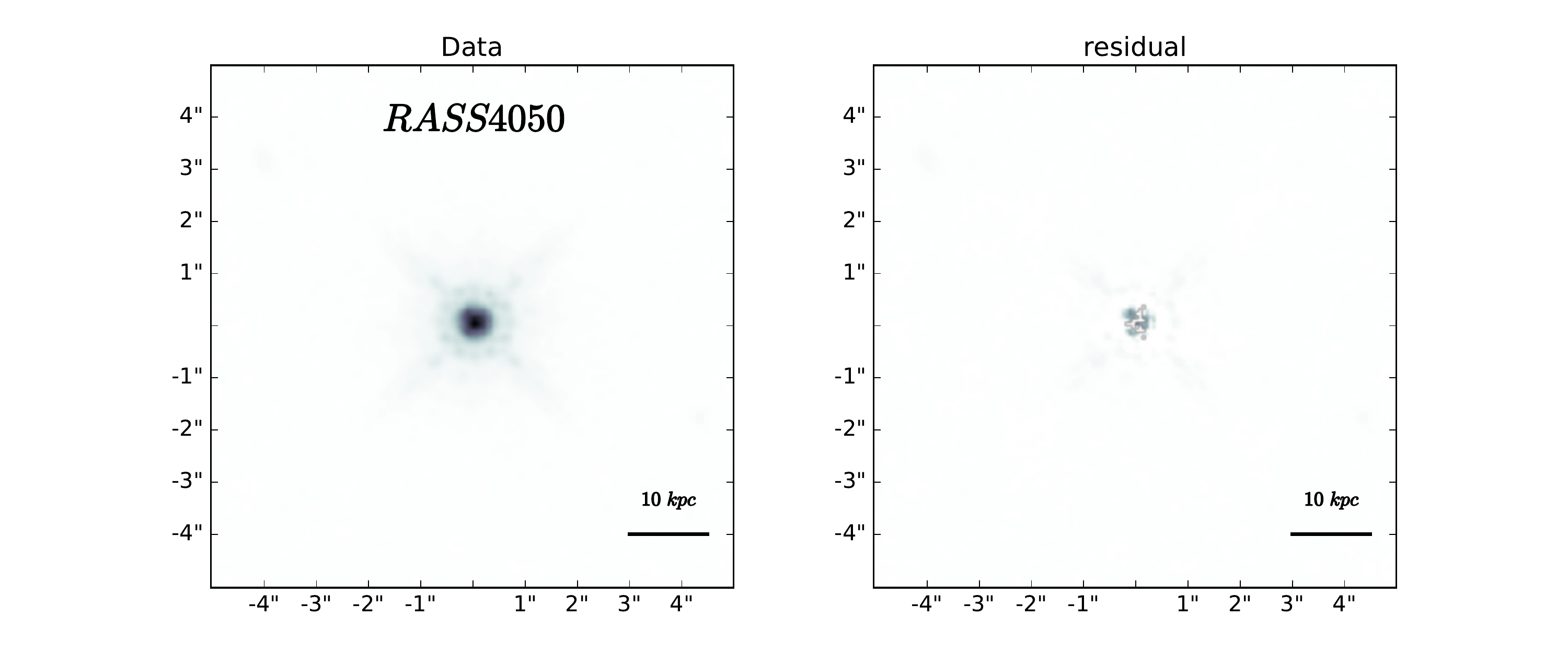}
\includegraphics[width=14cm]{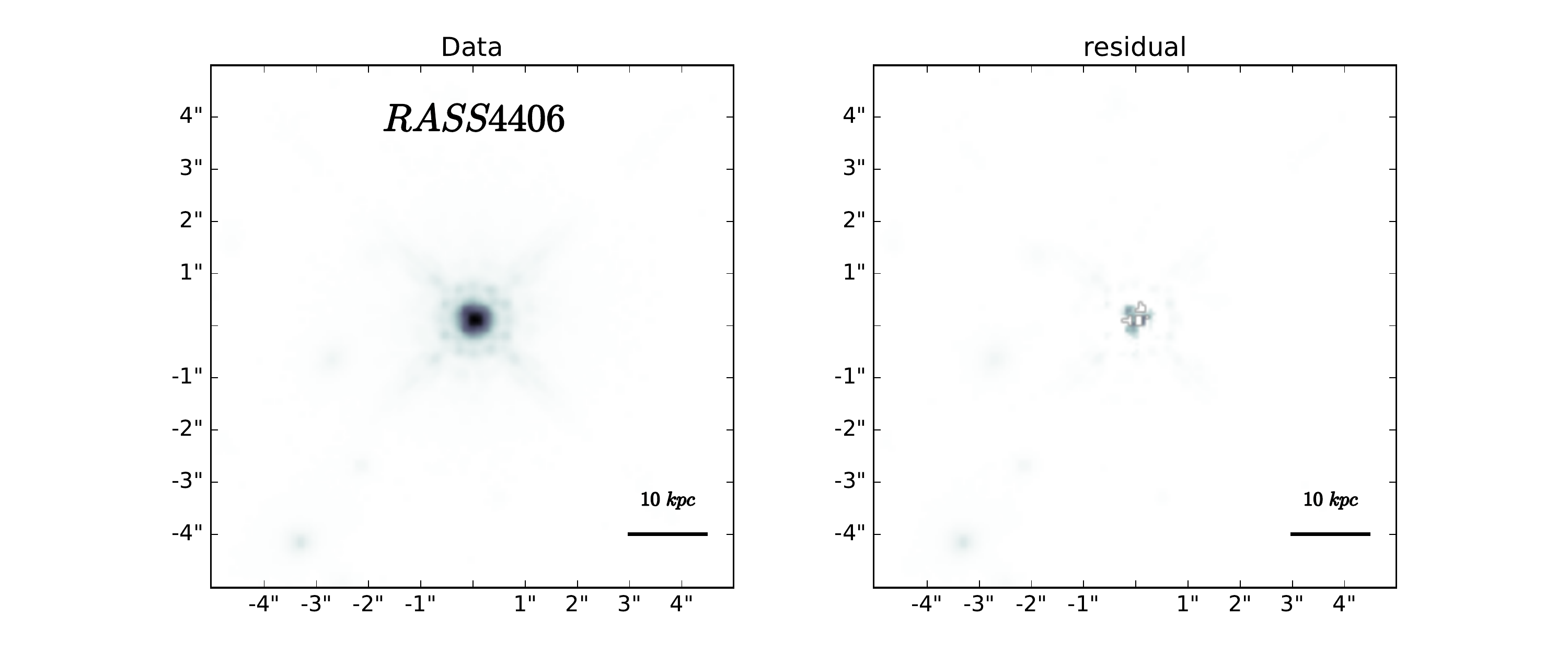}
\caption{\textsc{galfit} residuals for unresolved sources, left most column shows raw images, right column shows residual with point source subtracted. Individual cut-outs are 10\arcsec x 10\arcsec and are oriented as observed rather than North up and East left to show the similarity in PSF patterns.}
\label{F:resid_unres}
\end{center}
\end{figure*}

\begin{figure*}
\begin{center}
\includegraphics[width=14cm]{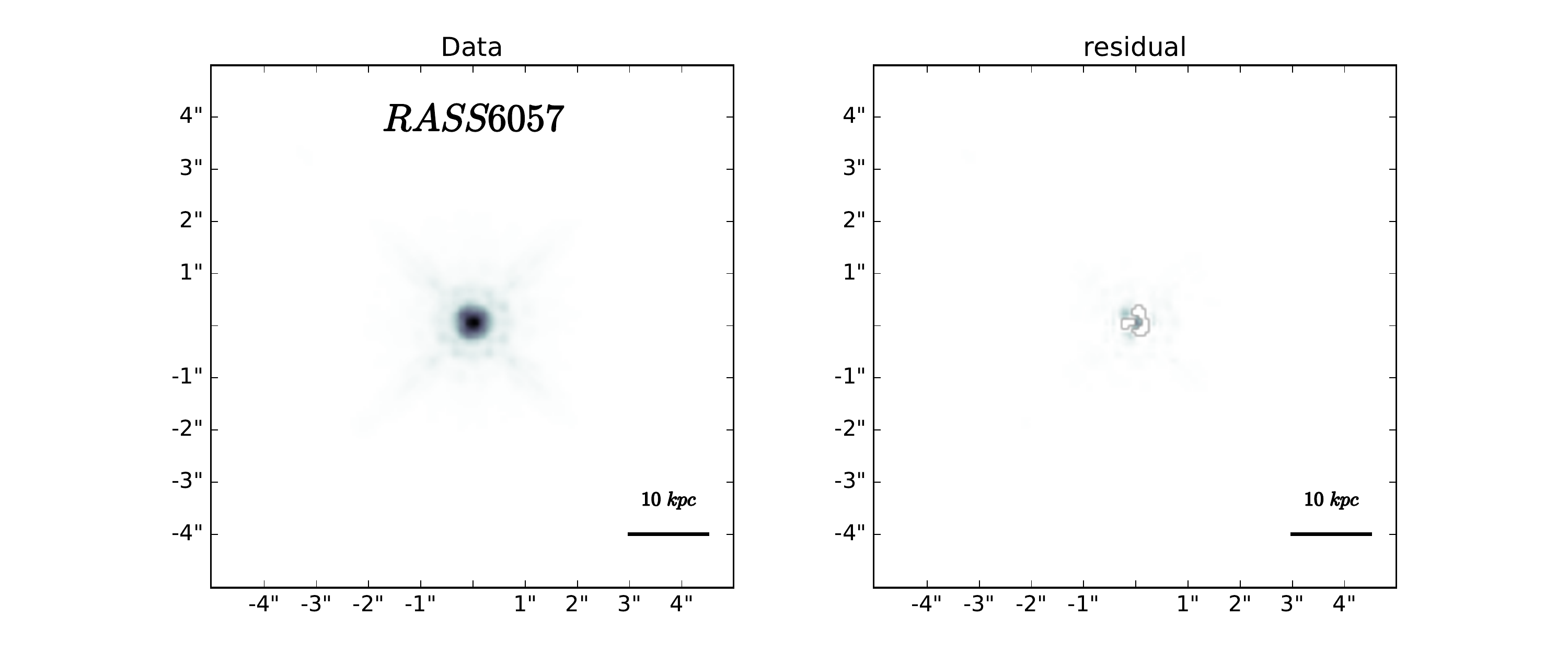}
\includegraphics[width=14cm]{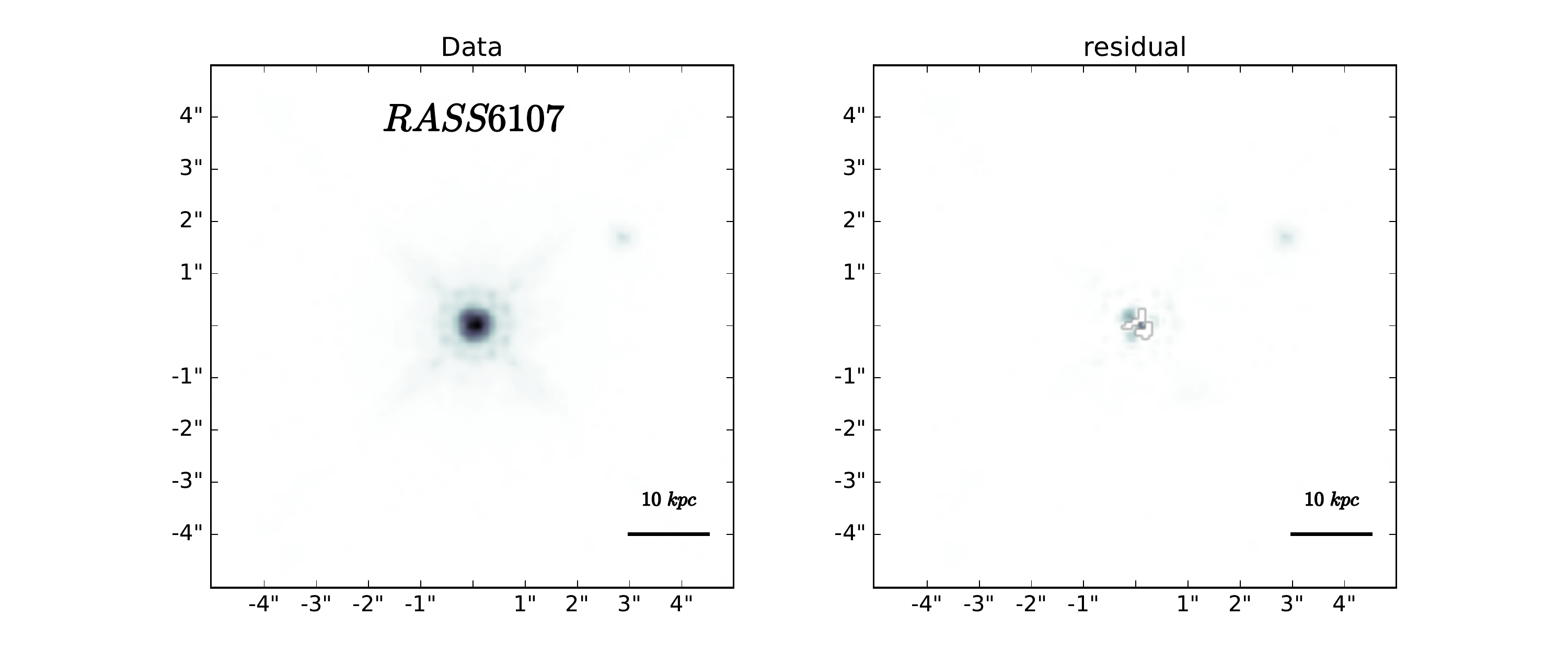}
\caption{\textsc{galfit} residuals for unresolved sources, continued, left most column shows raw images, right column shows residual with point source subtracted. Individual cut-outs are 10\arcsec x 10\arcsec and are oriented as observed rather than North up and East left to show the similarity in PSF patterns.}
\label{F:resid_unres}
\end{center}
\end{figure*}

\label{lastpage}

\end{document}